%% file: qrs-itcs.tex
\documentclass{sig-alternate}

\usepackage{amsmath,amssymb}
\usepackage{thmtools,thm-restate}

\usepackage[bookmarksnumbered,colorlinks,linkcolor=blue,citecolor=blue,urlcolor=blue,
  pdftitle={Quantum rejection sampling},
  pdfauthor={Maris Ozols, Martin Roetteler and Jeremie Roland}
]{hyperref}

\newcommand{\TikZorPDF}[2]{
#2
}

\TikZorPDF{
  \usepackage{tikz}
}{
  \usepackage{graphicx}
}

\newcounter{MYtempeqncnt}

\newcommand{\ket}[1]{|#1\rangle}
\newcommand{\bra}[1]{\langle#1|}
\newcommand{\braket}[2]{\langle#1|#2\rangle}
\newcommand{\set}[1]{\{#1\}}
\newcommand{\abs}[1]{|#1|}

\newcommand{\norm}[1]{\left\|#1\right\|}
\newcommand{\nix}[1]{}

\renewcommand{\H}[1]{H^{\otimes #1}} 
\newcommand{\mx}[1]{\begin{pmatrix}#1\end{pmatrix}} 

\newcommand{\vc}[1]{\boldsymbol{#1}} 
\newcommand{\bv}[1]{#1}

\newcommand{\R}{\mathbb{R}}
\newcommand{\C}{\mathbb{C}}
\newcommand{\F}{\mathbb{F}}
\newcommand{\E}{\mathbb{E}}
\renewcommand{\O}{\mathcal{O}}
\newcommand{\X}{\mathcal{X}}
\renewcommand{\P}{\mathcal{P}}
\renewcommand{\S}{\mathcal{S}}
\newcommand{\CC}{\mathcal{C}}
\newcommand{\EE}{\mathcal{E}}

\newcommand{\0}[1]{\ket{\bar{0}}_{#1}}
\newcommand{\Rot}{R_{\ve}}
\newcommand{\Rotk}{\Rot(k)}
\newcommand{\Ue}{U_{\ve}}

\renewcommand{\b}{\F_2}
\newcommand{\bb}[1]{\b^{#1}}
\newcommand{\oracle}[1]{O_{f_{#1}}}
\newcommand{\reflection}[1]{\mathrm{ref}_{#1}}

\newcommand{\e}{\varepsilon}
\newcommand{\ve}{\vc{\varepsilon}}
\newcommand{\pe}{p_{\ve}}
\newcommand{\qe}{q_{\ve}}

\newcommand{\pp}{p}

\newcommand{\psis }{\psi(\bv{s})}
\newcommand{\psifs}{\psi_{\smash{\hat{f}}}(\bv{s})}

\newcommand{\mc}[1]{\mathcal{#1}}
\newcommand{\fs}{f_{\bv{s}}}
\newcommand{\Vs}{V(\bv{s})}

\newcommand{\x}{\otimes}

\newcommand{\xp}[1]{^{\otimes #1}}
\newcommand{\ct}{^{\dagger}}
\newcommand{\tp}{^{\mathsf{T}}}

\newcommand{\gv}{\mathrm{I}_{f}(\bv{v})}
\newcommand{\gf}{\mathrm{I}_{f}}

\newcommand{\p}{P}
\newcommand{\q}{S}

\newcommand{\quantump}{\pi}
\newcommand{\quantums}{\sigma}
\newcommand{\statep}{\vc{\pi}}
\newcommand{\states}{\vc{\sigma}}
\newcommand{\statepxi}{\vc{\pi}^\xi}
\newcommand{\statesxi}{\vc{\sigma}^\xi}

\newcommand{\waterfilling}{\ve_{\vc{\quantump}\to\vc{\quantums}}^p}

\newcommand{\error}{\mathrm{error}}

\newcommand{\U}[1]{\mathrm{U}(#1)}

\newcommand{\sgeq}{\succeq}

\DeclareMathOperator{\diag}{diag}
\DeclareMathOperator{\Tr}{Tr}
\DeclareMathOperator{\spn}{span}
\DeclareMathOperator{\Herm}{Herm}
\DeclareMathOperator{\spec}{spec}

\renewcommand{\Re}{\operatorname{Re}}

\newcommand{\ie}{\textit{i.\,e.}}
\newcommand{\eg}{\textit{e.\,g.}}

\newcommand{\sampling}    {\textsc{Sampling}}
\newcommand{\qsampling}   {\textsc{QSampling}}
\newcommand{\qsamplingab} {\qsampling_{\vc{\quantump}\to\vc{\quantums}}}
\newcommand{\sqsampling}  {\textsc{SQSampling}}
\newcommand{\sqsamplingab}{\sqsampling_{\vc{\tau}}}
\newcommand{\bhsp}        {\textsc{BHSP}}
\newcommand{\qle}         {\textsc{QLinearEquations}_\kappa}
\newcommand{\qmm}         {\textsc{QMetropolisMove}_\CC}
\newcommand{\IE}          {\textsc{IndexErasure}}

\newcommand{\SQRS}{\mc{S}_{\text{QRS}}}
\newcommand{\AQRS}{\mc{A}_{\text{QRS}}}
\newcommand{\ASQRS}{\mc{A}_{\text{SQRS}}}

\newsavebox{\fmbox}
\newenvironment{fmpage}[1]
{\medskip\begin{lrbox}{\fmbox}\begin{minipage}{#1}}
{\end{minipage}\end{lrbox}\fbox{\usebox{\fmbox}}\medskip}

\newcommand{\algorithm}[1]{
\begin{center}
\begin{fmpage}{0.46\textwidth}
\small #1
\end{fmpage}
\end{center}}

\newcommand{\partitle}[1]{\paragraph{\bf #1}\nobreak\vspace{0.5em}}

\newtheorem{lemma}{Lemma}

\newdef{definition}{Definition}
\newcommand{\deftitle}[1]{} 

\begin{document}

\conferenceinfo{ITCS}{'12, January 08 - 10, 2012, Cambridge, MA, USA}
\CopyrightYear{2012} 
\crdata{978-1-4503-1115-1}  

\title{Quantum rejection sampling}

\numberofauthors{3}
\author{
\alignauthor
Maris Ozols\\
       \affaddr{NEC Laboratories America}\\
       \affaddr{and}\\
       \affaddr{University of Waterloo, IQC}\\
       \email{marozols@yahoo.com}
\alignauthor
Martin Roetteler\\
       \affaddr{NEC Laboratories America}\\
       \email{\mbox{mroetteler@nec-labs.com}}
\alignauthor
J\'er\'emie Roland\\
       \affaddr{NEC Laboratories America}\\
       \affaddr{and}\\
       \affaddr{QuIC - Ecole Polytechnique de Bruxelles}\\
       \affaddr{Universit\'e Libre de Bruxelles}\\
       \email{jroland@ulb.ac.be}
}

\maketitle

\begin{abstract}
Rejection sampling is a well-known method to sample from a target distribution, given the ability to sample from a given distribution. The method has been first formalized by von~Neumann (1951) and has many applications in classical computing. We define a quantum analogue of rejection sampling: given a black box producing a coherent superposition of (possibly unknown) quantum states with some amplitudes, the problem is to prepare a coherent superposition of the same states, albeit with different target amplitudes. The main result of this paper is a tight characterization of the query complexity of this quantum state generation problem. We exhibit an algorithm, which we call quantum rejection sampling, and analyze its cost using semidefinite programming. Our proof of a matching lower bound is based on the automorphism principle which allows to symmetrize any algorithm over the automorphism group of the problem. Our main technical innovation is an extension of the automorphism principle to continuous groups that arise for quantum state generation problems where the oracle encodes unknown quantum states, instead of just classical data. Furthermore, we illustrate how quantum rejection sampling may be used as a primitive in designing quantum algorithms, by providing three different applications. We first show that it was implicitly used in the quantum algorithm for linear systems of equations by Harrow, Hassidim and Lloyd. Secondly, we show that it can be used to speed up the main step in the quantum Metropolis sampling algorithm by Temme~\textit{et al.}. Finally, we derive a new quantum algorithm for the hidden shift problem of an arbitrary Boolean function and relate its query complexity to ``water-filling'' of the Fourier spectrum.
\end{abstract}

\section{Introduction}

We address the problem of preparing a desired target quantum state
into the memory of a quantum computer. It is of course unreasonable to try to find an efficient
quantum algorithm to achieve this for general quantum states. Indeed, if any state could be prepared efficiently such
difficult tasks as preparing witnesses for QMA-complete problems~\cite{KKR:2006}
could be solved efficiently, a task believed to be
impossible even if classical side-information about the quantum state
is provided~\cite{AD:2010}. On the other hand, many interesting
computational problems can be related to quantum state generation
problems that carry some additional {\em structure} which might be
exploited by an efficient algorithm.

Among the most tantalizing examples of problems that are
reducible to quantum state generation is the {\sc Graph
Isomorphism} problem~\cite{KST:93} which could be solved by
preparing the quantum state $\ket{\Gamma} = \frac{1}{\sqrt{n!}}
\sum_{\pi\in S_n} \ket{\Gamma^\pi}$, \ie, the uniform superposition
of all the permutations of a given graph $\Gamma$. By generating such
states for two given graphs, one could then use the standard SWAP-test~\cite{BCWW:2001}
to check whether the two states are equal or
orthogonal and therefore decide whether the graphs are isomorphic or
not. Furthermore, it is known that all problems in statistical zero
knowledge (SZK) can be reduced to instances of quantum state
generation~\cite{aharonov03}, along with gap-versions of closest
lattice vector problems~\cite{Regev:2004,AR:2005} and subgroup
membership problems for arbitrary subgroups of finite groups~\cite{Watrous:2000,Watrous2001,Friedl2003}.
Aside from brute-force attempts that try to solve quantum state preparation
by applying sequences of controlled
rotations (typically of exponential length) to fix the amplitudes of the target state one qubit at a
time, not much is known regarding approaches to tackle the quantum
state generation problem while exploiting inherent structure.  

In this regard, the only examples we are aware of are
(i)~a direct approach to generate states described by efficiently computable amplitudes~\cite{Grover2002},
(ii)~an approach via adiabatic computing~\cite{aharonov03} in which a sequence of local Hamiltonians has to be found such that the desired target state is the ground state of a final Hamiltonian and the overlap between intermediate ground states is large, and
(iii)~recent work on quantum analogues of classical annealing processes~\cite{BKS:2009,SBBK:2008} and of the Metropolis sampling procedure~\cite{TOV+:2009,YA:2010}.

Conversely, for some problems a {\em lower} bound on the complexity of
solving a corresponding quantum state generation problem would be
desirable, for instance to provide further evidence for the security
of quantum money schemes, see \eg~\cite{Aaronson:2009,FGH+:2010}.
Unfortunately, except for a recent result that generalizes the
adversary method to a particular case of quantum state generation problems (see~\cite{LMRSS11} and~\cite{AMRR:2011}), very little is known about lower bounds for quantum state generation problems in general.

\partitle{Rejection sampling}

The classical rejection sampling
method\footnote{It is also known as the accept/reject method or
  ``hit-and-miss'' sampling.} was introduced by
von~Neumann~\cite{vonNeumann51} to solve the following
\emph{resampling} problem: given the ability to sample according to some
probability distribution $\p$, one is asked to produce samples from
some other distribution $\q$. Conceptually, the method is extremely
simple and works as follows: let $\gamma \leq 1$ be the largest
scaling factor such that $\gamma\q$ lies under $\p$, formally,
$\gamma=\min_k (p_k/s_k)$. We accept a sample $k$ from $\p$ with
probability $\gamma s_k/p_k$, otherwise we reject it and repeat.  The
expected number $T$ of samples from $\p$ to produce one sample from $\q$
is then given by $T = 1/\gamma = \max_k(s_k/p_k)$. See also~\cite{devroye}
for further details and analysis of the method for
various special cases of $\p$ and $\q$. In a setting where access to the
distribution $\p$ is given by a black box, this has been proved to be
optimal by Letac~\cite{Letac75}.  The rejection sampling technique is
at the core of many randomized algorithms and has a wide range of
applications, ranging from
computer science to statistical physics, where it is used for
Monte Carlo simulations, the most well-known example being the
Metropolis algorithm~\cite{metropolis53}.

In the same way that quantum state preparation can be considered a
quantum analogue of classical sampling, it is natural to study a
quantum analogue of the classical resampling problem, \ie, the
problem of sampling from a distribution $\q$ given the ability to
sample from another distribution $\p$. We call this problem {\em
  quantum resampling} and define it to be the following analogue of
the classical resampling problem: given an oracle generating a
quantum state $\ket{\statepxi}= \sum_{k=1}^n \quantump_k \ket{\xi_k} \ket{k}$,
where the amplitudes $\quantump_k$ are known but the states $\ket{\xi_k}$
are unknown, the task is to prepare a target state
$\ket{\statesxi}=\sum_{k=1}^n \quantums_k \ket{\xi_k} \ket{k}$ with (potentially)
different amplitudes $\quantums_k$ but the same states $\ket{\xi_k}$. Note that while both the initial amplitudes $\quantump_k$ and the final amplitudes $\quantums_k$ are fixed and known, the fact that the states $\ket{\xi_k}$ are unknown makes the problem non-trivial.

\partitle{Related work}

The query complexity of quantum state generation problems was studied in~\cite{AMRR:2011}, where the adversary method was extended to this model and used to prove a tight lower bound on a specific quantum state generation problem called $\IE$. The adversary method was later extended to quantum state \emph{conversion} problems---where the goal is to convert an initial state into a target state--- and shown to be nearly tight in the bounded error case for any problem in this class, which includes as special cases state generation and the usual model of function evaluation~\cite{LMRSS11}. In all these cases, the input to the problem is classical, as the oracle encodes some hidden classical data. This is where the quantum resampling problem differs from those models, as in that case the oracle encodes unknown quantum states.

Grover~\cite{Grover00} considered a special case of the quantum resampling problem, where the initial state
$\ket{\statep}=\frac{1}{\sqrt{2^n}}\sum_x \ket{x}$
is a uniform superposition and one is given access to an oracle that for given input $x$ provides a classical description of $\quantums_x$, the amplitude in the target state
$\ket{\states}=\sum_x \quantums_x \ket{x}$.
We significantly extend the scope of Grover's technique by considering any initial superposition and improving the efficiency of the algorithm when only an approximate preparation of the target state is required.

Techniques related to quantum resampling have already been used implicitly as a useful primitive for building quantum algorithms. For instance, it was used in a paper by Harrow, Hassidim, and Lloyd~\cite{HHL:09} for the problem of solving a system of linear equations $A x = b$, where $A$ is an invertible matrix over the real or complex numbers whose entries are efficiently computable, and $b$ is a vector. The quantum algorithm in~\cite{HHL:09} solves the problem of preparing the state $\ket{x}=A^{-1}\ket{b}$ by applying the following three basic steps. First, use phase estimation on the state
$\ket{b}=\sum_k b_k\ket{\psi_k}$
to prepare the state
$\sum_k b_k\ket{\lambda_k}\ket{\psi_k}$,
where $\ket{\psi_k}$ and $\lambda_k$ denote the eigenstates and eigenvalues of $A$. Next, convert this state to
$\sum_k b_k\lambda_k^{-1}\ket{\lambda_k}\ket{\psi_k}$.
Finally, undo the phase estimation to obtain the target state
$A^{-1}\ket{b}=\sum_k b_k\lambda_k^{-1}\ket{\psi_k}$.
The second step of this procedure performs transformation
$\sum_k b_k\ket{\lambda_k}\ket{\psi_k} \mapsto\sum_k b_k\lambda_k^{-1}\ket{\lambda_k}\ket{\psi_k}$
which can be seen as an instance of quantum resampling. Note that other works, such as~\cite{Childs2009a,Sheridan2009}, have used a similar technique---i.e., using phase estimation to simulate some function of an operator---to apply a unitary on an unknown quantum state, rather than preparing one particular sate.

The quantum Metropolis sampling algorithm has been proposed by Temme~\textit{et al.}~\cite{TOV+:2009} to solve the problem of preparing the thermal state of a quantum Hamiltonian. As it is heavily inspired by the classical Metropolis algorithm, the main step uses an accept/reject rule on random moves between eigenstates of the Hamiltonian. The main complication comes from reverting rejected moves, as the no-cloning principle prevents from keeping a copy of the previous eigenstate. We will show that this step actually reduces to a quantum resampling problem, and that quantum rejection sampling leads to an alternative solution which also provides a speed-up over the technique proposed in~\cite{TOV+:2009}.

Finally, another type of quantum resampling problem has been considered in a paper by Kitaev and Webb~\cite{KW:2009} in which the task is to prepare a superposition of basis states with Gaussian-distributed weights along a low-dimensional strip inside a high-dimensional space. Authors solve this problem by applying a sequence of lattice transformation and phase estimation steps.

For us, another important case in point are hidden shift problems over an abelian group $A$. Here it is easy to prepare a quantum state of the form $\ket{\statepxi} = \sum_{\bv{w}\in A} \hat{f}(\bv{w}) \chi_{\bv{w}}(s)  \ket{\bv{w}}$, where $\chi_{\bv{w}}$ denotes the characters of $A$ and $\hat{f}$ denotes the Fourier transform of $f$ (see e.\,g., \cite{vDHI:2003,Ivanyos:2008,Roetteler:2010}). If we could eliminate the Fourier coefficients $\hat{f}(\bv{w})$ from state $\ket{\statepxi}$, we would obtain a state $\ket{\statesxi} = |A|^{-1/2} \sum_{\bv{w}\in A} \chi_{\bv{w}}(s) \ket{\bv{w}}$ from which the hidden shift $\bv{s}$ can be easily recovered by applying another Fourier transform. Note that in this case the states $\ket{\xi_k}$ are actually just the complex phases $\chi_{\bv{w}}(s)$. We will discuss an application of our general framework to the special case of hidden shift problems in Sect.~\ref{sect:hiddenshift}.

\partitle{Our results}

\begin{figure}[th]
  
  \centering
  \TikZorPDF{
    \input{fig-model.tex}
  }{
    \includegraphics[width=0.35\textwidth]{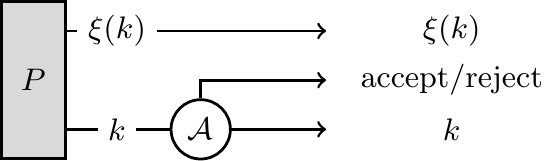}
  }

  \caption{Classical rejection sampling: A black box produces samples $k$ according to a known probability distribution $\p$ and accompanied by some unknown classical data $\xi(k)$. The algorithm $\mathcal{A}$ either accepts or rejects these samples, so that accepted samples are distributed according to a target distribution $\q$.}
  \label{fig:CRS}
\end{figure}

We denote the classical resampling problem mentioned above by $\sampling_{\p \to \q}$, where $\p$ and $\q$ are probability distributions on the set $[n]$. Note that in its simplest form, this problem is not meaningful in the context of query complexity (indeed, if distribution $\q$ is known to begin with, there is no need to use the ability to sample from $\p$). However, there is a natural modification of the problem, that actually models realistic applications, which does not suffer from this limitation. In this version of the problem, there is a function $\xi$ that deterministically associates some unknown data with each sample, and the problem is to sample pairs $(k,\xi(k))$, where $k$ follows the target distribution. Formally, the problem is therefore defined as follows: given oracle access to a black box producing pairs $(k,\xi(k))\in [n]\times[d]$ such that $k$ is distributed according to $\p$, where $\xi:[n]\to[d]$ is an unknown function, the problem is to produce a sample $(k,\xi(k))$ such that $k$ is distributed according to $\q$. Note that in this model it is not possible to produce the required samples without using the access to the oracle that produces the samples from $\p$, and the algorithm is therefore restricted to act as in Fig.~\ref{fig:CRS}.

The problem studied in this article is a quantum analogue of
$\sampling_{\p \to \q}$, where probability distributions are replaced by
quantum superpositions.  More precisely, let $\vc{\quantump},\vc{\quantums}\in\R^n$ be such that
  $\norm{\vc{\quantump}}_2=\norm{\vc{\quantums}}_2=1$ and
  $\quantump_k,\quantums_k\geq 0$ for all $k\in [n]$. Let $O$ be a unitary
  that acts on a default state $\0{dn} \in \C^d \x \C^n$ as
  $O:\0{dn}\mapsto\ket{\statepxi} := \sum_{k=1}^n \quantump_k \ket{\xi_k}
  \ket{k},$ where $\ket{\xi_k} \in \C^d$ are some fixed unknown
  normalized quantum states. Given oracle access to unitary black
  boxes $O,O^\dagger$, the $\qsamplingab$ problem is
  to prepare the state
  $\ket{\statesxi} := \sum_{k=1}^n \quantums_k \ket{\xi_k} \ket{k}$.
Note that a special case of this problem is the scenario $d = 1$, when
$\xi_k \in \C$ are just unknown phases (complex numbers of absolute
value $1$).

The main result of this article is a tight characterization of the
query complexity of $\qsamplingab$ for any success probability $p$ (the vector $\waterfilling$, as well as the probabilities $p_{\min},p_{\max}$, will be defined in Sect.~\ref{sect:problem}, intuitively, the vector $\waterfilling$ characterizes the amplitudes of the final state prepared by the best algorithm having success probability $p$):
\begin{restatable}{theorem}{thmqsampling}
\label{thm:qsampling}
For $p \in [p_{\min}, p_{\max}]$, the quantum query complexity of $\qsamplingab$ with success probability $p$ is $Q_{1-p}(\qsamplingab)=\Theta({1}/{\norm{\waterfilling}_2})$.
For $p\leq p_{\min}$, the query complexity is 1, and for $p> p_{\max}$, it is infinite.
\end{restatable}

\noindent Let us note that when $p=p_{\max}=1$, the query complexity reduces to $\max_k(\quantums_k/\quantump_k)$ in analogy with the classical case, except that amplitudes replace probabilities. The lower bound comes from an extension of the automorphism principle (originally introduced in the context of the adversary method~\cite{AMRR:2011,NegativeWeights}) to our framework of quantum state generation problems with quantum oracles. The upper bounds follows from an algorithm based on amplitude amplification that we call quantum rejection sampling. We also show that a modification of this algorithm can also solve a quantum state conversion problem, which we call strong quantum resampling ($\sqsampling$).

Next, we illustrate the technique by providing different applications. We first show that the main steps in two recent algorithms, namely the quantum algorithm for solving linear systems of equations~\cite{HHL:09} and the quantum Metropolis sampling algorithm~\cite{TOV+:2009}, consists in solving quantum state conversion problems which we call $\qle$ and $\qmm$. We then observe that these problems reduce to $\sqsampling$, and can therefore be solved using quantum rejection sampling.
\begin{restatable}{theorem}{thmlinearequations}
\label{thm:linear-equations}
For any $\tilde{\kappa} \in [1,\kappa]$, there is a quantum algorithm that solves $\qle$ with success probability $p = (\vc{w}\tp \cdot \tilde{\vc{w}}) / (\norm{\vc{w}}_2 \cdot \norm{\tilde{\vc{w}}}_2)$ using an expected number of queries $O(\tilde{\kappa}/\norm{\tilde{\vc{w}}}_2)$, where $w_j := b_j / \lambda_j$, $\tilde{w}_j := b_j / \tilde{\lambda}_j$, and $\tilde{\lambda}_j := \max \set{\tilde{\kappa}^{-1}, \lambda_j}$.
\end{restatable}
\begin{restatable}{theorem}{thmmetropolis}
\label{thm:metropolis}
There is a quantum algorithm that solves $\qmm$ with success probability 1 using an expected number of queries $O(1/\|\vc{w}^{(i)}\|_2)$.
\end{restatable}
Let us note that while the quantum algorithm for linear systems of equations was indeed using this technique implicitly, this was not the case for quantum Metropolis sampling, where quantum rejection sampling provides some speed-up over the original algorithm.

Our final result is an application of quantum rejection sampling to the Boolean
hidden shift problem $\bhsp_f$, defined as follows. Let $f:
\bb{n} \to \b$ be a Boolean function, which is assumed to be
completely known. Furthermore, we are given oracle access to
\emph{shifted function} $\fs(\bv{x}) := f(\bv{x} + \bv{s})$ for some
unknown bit string $\bv{s} \in \bb{n}$, with the promise that there
exists $\bv{x}$ such that $f(\bv{x} + \bv{s})\neq f(\bv{x})$. The task 
is to find the bit string $\bv{s}$.

We show that we can solve this problem by solving the
$\qsamplingab$ problem for $\vc{\quantump}$
corresponding to the Fourier spectrum of $f$, and $\vc{\quantums}$ being a
uniformly flat vector. This leads to the following upper bound which
expresses the complexity of our quantum algorithm for the Boolean
hidden shift problem in terms of a vector $\waterfilling$ (defined in
Sect.~\ref{sect:SDP}) that can be thought of as a ``water-filling'' of
the Fourier spectrum of $f$: 

\begin{restatable}{theorem}{thmBHSP}
Let $f$ be a Boolean function and $\hat{f}$ be its Fourier transform. For any $p$, we have $Q_{1-p}(\bhsp_f) = O(1/\norm{\waterfilling}_2)$, where components of $\vc{\quantump}$ and $\vc{\quantums}$ are given by $\quantump_{\bv{w}}=\abs{\hat{f}_{\bv{w}}}$ and $\quantums_{\bv{w}}=1/\sqrt{2^n}$ for $\bv{w} \in \bb{n}$.
\label{thm:BHSP}
\end{restatable}

As special cases of this theorem we obtain the quantum algorithms for
hidden shift problem for delta functions, which leads to the Grover
search algorithm~\cite{Grover:96}, and for bent functions, which are
functions that have perfectly flat absolute values in their Fourier
spectrum~\cite{Roetteler:2010}. In general, the complexity of the algorithm is limited by the smallest Fourier coefficient of the function. By ignoring small Fourier coefficients, one can decrease the complexity of the algorithm, at the cost of a lower success probability. The final success probability can nevertheless be amplified using repetitions and by constructing a procedure to check a candidate shift, which leads to the following theorem.
\begin{restatable}{theorem}{thmBHSPboost}
\label{thm:BHSP2}
Let $f$ be a Boolean function and $\hat{f}$ be its Fourier transform. Moreover, let $p,\gamma\in[0,1]$ be such that $\norm{\hat{\ve}}_1 = \sqrt{2^n p}$, where $\e_w=\min \set{\abs{\hat{f}_w}, \gamma/\sqrt{2^n}}$. Then, for any $\delta > 0$, we have $Q_{\delta}(\bhsp_f) = O\bigl( \frac{1}{\sqrt{p}} (1/\norm{\ve}_2 + 1/\sqrt{\gf}) \bigr)$.
\end{restatable}

\section{Definition of the problem} \label{sect:problem}

In this section, we define different notions related to the quantum resampling problem. It is important to note that this problem goes beyond the usual model of quantum query complexity, where the goal is to compute a function depending on some unknown classical data that can be accessed via an oracle (see~\cite{BuhrmanDeWolf02querysurvey} for a complete survey). In the usual model, the algorithm is therefore quantum but both its input and output are classical. A first extension of this model is the case were the output is also quantum, that is, the goal is to generate a target quantum state depending on some unknown classical data hidden by the oracle. The quantum adversary method has recently been extended to this model by~\cite{AMRR:2011}, where is was used to characterize the query complexity if a quantum state generation problem called $\textsc{IndexErasure}$. In both the usual model and this extension, the oracle acts as $O_x:\ket{b}\ket{i}\mapsto\ket{b+x_i}\ket{i}$, where $x$ is the hidden classical data. However, the quantum resampling problem corresponds to another extension of these models, where the input is also quantum, in the sense that the oracle hides unknown quantum states instead of classical data. Let us now define this extended model more precisely.
\begin{definition}\deftitle{Quantum state generation problem}
Let $\O := \set{O_x : x \in \X}$ and $\Psi := \set{ \ket{\psi_x} : x \in \X}$, respectively, be sets of quantum oracles (\ie, unitaries) and target quantum states labeled by elements of some set $\X$. Then $\P:=(\O, \Psi, \X)$ describes the following \emph{quantum state generation problem:} given an oracle $O_x$ for some unknown value of $x \in \X$, prepare a state
\begin{equation*}
 \ket{\psi}=\sqrt{p}\ket{\psi_x}\0{}+\ket{\error_x},
\end{equation*}
where $p$ is the desired success probability, $\0{}$ is a normalized standard state for some workspace and $\ket{\error_x}$ is an arbitrary error state with norm at most $\sqrt{1-p}$. The quantum query complexity of $\P$ is the minimum number of queries to $O_x$ or $O_x^\dagger$ necessary to solve $\P$ with success probability $p$ and will be denoted by $Q_{1-p}(\P)$.
\end{definition}

Intuitively, we want the final state $\ket{\psi}$ to have a component of length at least $\sqrt{p}$ in the direction of $\ket{\psi_x}\0{}$. We can restate the condition $\norm{\ket{\error_x}}_2 \leq \sqrt{1-p}$ as follows:
\begin{equation*}
1 - p \geq \norm{\ket{\psi} - \sqrt{p} \ket{\psi_x} \0{}}_2^2 = 1 + p - 2 \Re \bigl[ \bra{\psi} \cdot \sqrt{p} \ket{\psi_x} \0{} \bigr],
\end{equation*}
or equivalently:
\begin{equation}
  \Re \bigl[ \bra{\psi} \cdot \ket{\psi_x} \0{} \bigr] \geq \sqrt{\pp}.
  \label{eq:Closeness}
\end{equation}

Note that the main difference of the above definition with the usual model of quantum query complexity, and its extension to quantum state generation in~\cite{AMRR:2011}, is that the oracle is not restricted to act as $O_x:\ket{b}\ket{i}\mapsto\ket{b+x_i}\ket{i}$.

We now formally define $\qsamplingab$ as a special case of quantum state generation problem. Throughout this article, we fix positive integers $d,n$ and we assume that $\vc{\quantump},\vc{\quantums}\in\R^n$ are vectors such that $\norm{\vc{\quantump}}_2=\norm{\vc{\quantums}}_2=1$ and
 $\quantump_k,\quantums_k\geq 0$ for all $k\in [n]$. We also use the notation $\ket{\statep}:=\sum_{k=1}^n \quantump_k \ket{k}$ and $\ket{\states}:=\sum_{k=1}^n \quantums_k \ket{k}$. For simplicity, we assume that $\quantums_k>0$ for all $k\in[n]$, but the general case can easily be obtained by taking the limit $\quantums_k\to 0$.

Let $\xi = (\ket{\xi_k} \in \C^d : k \in [n])$ be an ordered list of normalized quantum states. Then any unitary that maps a default state $\0{dn}$ to $\ket{\statepxi} := \sum_{k=1}^n \quantump_k \ket{\xi_k} \ket{k}$ is a valid oracle for $\qsamplingab$. Therefore, we will label valid oracles by a pair $(\xi,u)$, where $\xi$ denotes the states hidden by the oracle, and $u$ defines how the oracle acts on states that are orthogonal to $\0{dn}$.
More explicitly, we fix a default oracle $O \in \U{dn}$ as a unitary that acts on $\0{dn}$ as $O \0{dn} = \0{d} \ket{\statep}$, and arbitrarily on the orthogonal complement. We then use $O$ as a reference point to define the remaining oracles:
\begin{equation}
  O_{\xi,u} := V_{\xi} \, O \, u,
  \label{eq:oracles}
\end{equation}
where $u \in \U{dn}$ is a unitary such that $u\0{dn}=\0{dn}$ and $V_\xi$ is a unitary that acts on $\0{d}\ket{k}$ as $V_\xi\0{d}\ket{k}=\ket{\xi_k}\ket{k}$ for any $k\in [n]$, and arbitrarily on the orthogonal complement of these states, so that $O_{\xi,u} \0{dn} = V_{\xi} \, O \0{dn} = V_{\xi} \sum_{k=1}^n \quantump_k \0{d} \ket{k} = \sum_{k=1}^n \quantump_k \ket{\xi_k} \ket{k} = \ket{\statepxi}$ (note that how $O$ and $V_\xi$ are defined on the orthogonal complements is irrelevant as it only affects the exact labeling, but not the set of valid oracles).

\begin{definition}\deftitle{Quantum resampling problem}
The \emph{quantum resampling problem}, denoted by $\qsamplingab$, is an instance of quantum state generation problem $(\O, \Psi, \X)$ with
\begin{align*}
  \X &:= \set{(\xi,u) : \xi = (\ket{\xi_1}, \dotsc, \ket{\xi_n}) \in (\C^d)^n, u \in S}, \\
   S &:= \set{u\in\U{dn}:u\0{dn}=\0{dn}}\cong \U{dn-1}.
\end{align*}
Oracles in $\O$ that are hiding the states $\ket{\statepxi}$ are defined according to Eq.~(\ref{eq:oracles}) and the corresponding target states are defined by $\ket{\statesxi} := V_{\xi}\0{d} \ket{\states} = \sum_{k=1}^n \quantums_k\ket{\xi_k} \ket{k}$.
\end{definition}
Let us note that the target states only depend on the index $\xi$, and not $u$. Moreover, note that amplitudes $\quantump_k$ and $\quantums_k$ can be assumed to be real and positive without loss of generality, as any phase can be corrected using a controlled-phase gate, which does not require any oracle call since $\vc{\quantump}$ and $\vc{\quantums}$ are fixed and known.

\begin{figure}[th]
  
  \centering
  \TikZorPDF{
    \input{fig-problems.tex}
  }{
    \includegraphics[width=0.45\textwidth]{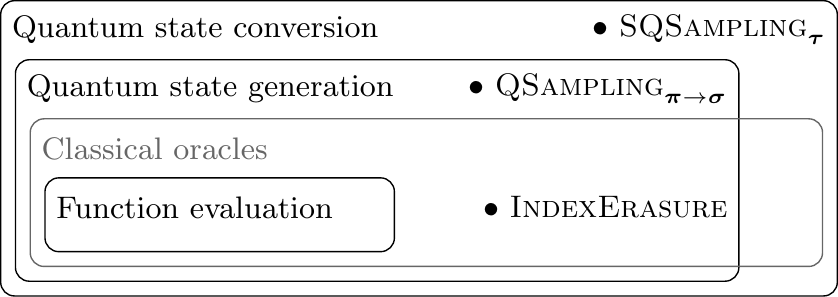}
  }

  \caption{Comparison of different classes of problems in quantum query complexity. The case of function evaluation has been extensively studied in the literature. The extension to quantum state generation with classical oracles, as well as the problem \textnormal{$\IE$} which belongs to that class, have been studied in~\cite{AMRR:2011}. The adversary method has been extended to the case of quantum state conversion with classical oracles in~\cite{LMRSS11}. The problems \textnormal{$\qsamplingab$} and \textnormal{$\sqsamplingab$} studied in this article use quantum oracles and therefore belong to yet another extension of the quantum query complexity model.}
  \label{fig:problems}
\end{figure}

In~\cite{LMRSS11}, Lee~\textit{et al.} have proposed another extension of the query complexity model for quantum state generation of~\cite{AMRR:2011} by considering quantum state conversion, where the problem is now to convert a given quantum state into another quantum state, instead of generating a target quantum state from scratch. They have extended the adversary method to this model and showed that it is approximately tight in the bounded-error case for any quantum state conversion problem with a classical oracle. Here, we define a model that combines both extensions (from classical to quantum oracles as well as from state generation to state conversion), hence subsuming all previous models (see Fig.~\ref{fig:problems}).
\begin{definition}\deftitle{Quantum state conversion problem}
Let $\O := \set{O_x : x \in \X}$, $\Phi := \set{ \ket{\phi_x} : x \in \X}$ and $\Psi := \set{ \ket{\psi_x} : x \in \X}$, respectively, be sets of quantum oracles (\ie, unitaries), initial quantum states and target quantum states labeled by elements of some set $\X$. Then $\P:=(\O, \Phi, \Psi, \X)$ describes the following \emph{quantum state conversion problem:} given an oracle $O_x$ for some unknown value of $x \in \X$ and a copy of the corresponding initial state $\ket{\phi_x}$, prepare a state
\begin{equation}
 \ket{\psi}=\sqrt{p}\ket{\psi_x}\0{}+\ket{\error_x},
\end{equation}
where $p$ is the desired success probability, $\0{}$ is a normalized standard state for some workspace and $\ket{\error_x}$ is an arbitrary error state with norm at most $\sqrt{1-p}$. Again, $Q_{1-p}(\P)$ will denote the quantum query complexity of $\P$.
\end{definition}

We also define a strong version of the quantum resampling problem, which is a special case of the state conversion problem. Compared to the original resampling problem, it is made harder due to two modifications. First, instead of being given access to an oracle that maps $\0{dn}$ to $\ket{\statepxi}$, we are only provided with one copy of $\ket{\statepxi}$ and an oracle that reflects through it, making this a quantum state conversion problem. The second extension assumes that we only know the ratios between the amplitudes $\quantump_k$ and $\quantums_k$ for each $k$, instead of the amplitudes themselves. More precisely, instead of vectors $\vc{\quantump},\vc{\quantums}\in\R^n$ specifying the initial and target amplitudes, we fix a single vector $\vc{\tau}\in\R^n$ such that $\tau_k \geq 0$ and $\max_k \tau_k = 1$, specifying the ratios between those amplitudes. Let us now formally define the stronger version of the quantum resampling problem (this definition is motivated by the applications that will be presented in Sect.~\ref{sect:linear} and~\ref{sect:metropolis}).
 
\begin{definition}\deftitle{Strong quantum resampling problem}
Let $P := \set{\vc{\quantump} \in \R^n : \norm{\vc{\quantump}}_2 = 1, \forall k: \quantump_k > 0}$. The \emph{strong quantum resampling problem $\sqsamplingab$} is a quantum state conversion problem $(\O, \Phi, \Psi, \X)$, where $\X := \set{(\xi,\vc{\quantump}) : \xi = (\ket{\xi_1}, \dotsc, \ket{\xi_n}) \in (\C^d)^n, \vc{\quantump} \in P}$, oracles in $\O$ are defined by $O_{\xi,\vc{\quantump}}:=\reflection{\ket{\statepxi}} = I - 2 \ket{\statepxi} \bra{\statepxi}$ with the corresponding initial and target states being $\ket{\statepxi}$ and $\ket{\statesxi} = \sum_{k=1}^n \quantums_k\ket{\xi_k} \ket{k}$, respectively, where $\vc{\quantums} := \vc{\quantump} \circ \vc{\tau} / \norm{\vc{\quantump} \circ \vc{\tau}}_2$ so that $\quantums_k / \quantump_k \propto \tau_k$.
\label{def:SQSampling}
\end{definition}

The relationship between different classes of query complexity problems introduced above, and strong and regular quantum rejection sampling as special instances of them are summarized in Fig.~\ref{fig:problems}. Our main result is that the quantum query complexities of $\qsamplingab$ and $\sqsamplingab$ for any success probability $p$ depend on a vector $\waterfilling$ defined as follows.

\begin{definition}\label{def:water-filling}
For any $\vc{\quantump},\vc{\quantums}$, let us define the following quantities
\begin{align*}
  p_{\min} &:= (\vc{\quantums}\tp \cdot \vc{\quantump})^2, &
  \gamma_{\min} &:= \min_{k:\quantump_k>0} (\pi_k/\sigma_k), \\
  p_{\max} &:= \sum_{k:\quantump_k>0} \quantums_k^2, &
  \gamma_{\max} &:= \max_{k} (\pi_k/\sigma_k).
\end{align*}
For any $\gamma\in[\gamma_{\min},\gamma_{\max}]$, let us define a vector $\ve(\gamma)$ and a scalar $p(\gamma)$ by
\begin{align*}
 \e_k(\gamma)&:=\min \set{\quantump_k, \gamma \quantums_k}, &
 p(\gamma)&:=\biggl(\frac{\vc{\quantums}\tp\cdot\ve(\gamma)}{\norm{\ve(\gamma)}_2}\biggr)^2.
\end{align*}
For $p \in [p_{\min}, p_{\max}]$, let $\bar{\gamma}\in[\gamma_{\min},\gamma_{\max}]$ be such that $p(\bar{\gamma})=p$ and define a vector $\waterfilling := \ve(\bar{\gamma})$.
\end{definition}

To see that $\waterfilling$ is well-defined, note that $\norm{\ve(\gamma)}_2$ is monotonically increasing with $\gamma$, while $p(\gamma)$ is monotonically decreasing with $\gamma$. More precisely, for $\gamma=\gamma_{\min}$, the vector $\ve(\gamma)$ has components $\e_k(\gamma) = \gamma \quantums_k$ if $\pi_k \neq 0$ or zero otherwise, and $p(\gamma)=p_{\max}$. For $\gamma=\gamma_{\max}$, we have $\ve(\gamma)=\vc{\quantump}$ and $p(\gamma)=p_{\min}$. Between these extreme cases, $p(\gamma)$ interpolates from $p_{\max}$ to $p_{\min}$, which means that for any $p\in[p_{\min}, p_{\max}]$, there exists a value $\bar{\gamma}$ such that $p(\bar{\gamma})=p$, which uniquely defines $\waterfilling$.

Intuitively, $\ve(\gamma)$ may be interpreted as a ``water-filling'' vector, where $\gamma$ defines the water level, and $\quantump_k$ defines the height of ``tank'' number $k$. As $\gamma$ increases from $0$ to $\gamma_{\min}$, we have $\e_k(\gamma) = \gamma \quantums_k$, meaning that all tanks are progressively filled proportionally to $\gamma$. When $\gamma>\gamma_{\min}$, some tanks are filled ($\gamma \quantums_k>\quantump_k$) and cannot hold more water, while others continue to get filled.

\section{Query complexity of quantum resampling} \label{sect:SDP}

Let us first show that $\waterfilling$ defines an optimal feasible point of a certain semidefinite program (SDP). Afterwards we will show that the same SDP characterizes the quantum query complexity of $\qsamplingab$.

\begin{restatable}{lemma}{lemSDPsolution}\label{lem:sdp-solution}
Let $p\in[p_{\min}, p_{\max}]$, and $\ve=\waterfilling$. Then, the following SDP
\begin{equation}
\begin{array}{rl}
  \max_{M \sgeq 0} \Tr M
  \quad \text{s.t.} \quad
  & \forall k: \quantump_k^2 \geq M_{kk}, \\
  & \Tr \bigl[ (\vc{\quantums} \cdot \vc{\quantums}\tp - \pp I) M \bigr] \geq 0.
\end{array}
\label{eq:Primal}
\end{equation}
has optimal value $\norm{\ve}_2^2$, which is achieved by the rank-1 matrix $M=\ve\cdot\ve\tp$.
\end{restatable}

\begin{proof}[sketch]
We first show that $M=\ve\cdot\ve\tp$, where $\ve=\waterfilling$, satisfies the constraints in SDP~(\ref{lem:sdp-solution}) and therefore constitutes a feasible point. Therefore, the optimal value of~(\ref{lem:sdp-solution}) is at least $\Tr M=\norm{\ve}_2^2$. Secondly, we dualize the SDP, and provide a dual-feasible point achieving the same objective value. The fact that this objective value is feasible for both the primal and the dual then implies that this is the optimal value.
The details of the proof are given in Appendix~\ref{apx:SDP}.
\qed
\end{proof}

Next, let us prove that SDP~(\ref{eq:Primal}) provides a lower bound for the $\qsamplingab^\pp$ problem. In Sect.~\ref{sect:Algorithm}, we will also show that this lower bound is tight by providing an explicit algorithm.

Let us emphasize the fact that the lower bound cannot be obtained from standard methods such as the adversary method~\cite{Amb00,NegativeWeights} (which has recently been proved to be tight for evaluating functions~\cite{Reichardt:2009,ReichardtReflections,LMRSS11}), nor from its extension to quantum state generation problems~\cite{AMRR:2011,LMRSS11}, because in this case the oracle is also quantum, in the sense that it encodes some unknown quantum state rather than some unknown classical data. To prove lower bounds it is useful to exploit possible symmetries of the problem. We extend the notion of automorphism group~\cite{AMRR:2011,NegativeWeights} to our framework of quantum state generation problems:
\begin{definition}\deftitle{Automorphism group}
We say that $G$ is the \emph{automorphism group} of problem $(\O, \Psi, \X)$ if:
\begin{enumerate}
  \item $G$ acts on $\X$ (and thus implicitly also on $\O$ as $g: O_x \mapsto O_{g(x)}$).
  \item For any $g \in G$ there is a unitary $U_g$ such that $U_g \ket{\psi_x} = \ket{\psi_{g(x)}}$ for all $x \in \X$.
  \item For any given $g \in G$ it is possible to simulate the oracles $O_{g(x)}$ for all $x \in \X$, using only a constant number of queries to the black box $O_x$.
\end{enumerate}
\end{definition}
While for the standard model of quantum query complexity, where the oracle encodes some unknown classical data, the automorphism group is restricted to be a permutation group and is therefore always finite, in this more general framework the automorphism group can be continuous. For example, in the case of $\qsamplingab^\pp$ it will involve the unitary group.
Taking such symmetries into account might significantly simplify the analysis of algorithms for the corresponding problem and is the key to prove our lower bound.

\newcommand{\g }[3]{     \gamma _{#1}(#2,#3)}
\newcommand{\gb}[1]{\bar{\gamma}_{#1}}

\newcommand{\dm}[1]{\;d\mu(#1)}
\newcommand{\twirling      }[1]{\twirlingwithcoefficient{\frac{1}{\sqrt{2^n}}}{#1}}
\newcommand{\twirlingsquare}[1]{\twirlingwithcoefficient{\frac{1}{      2^n }}{#1}}
\newcommand{\twirlingwithcoefficient}[2]{#1 \sum_{\bv{y} \in \bb{n}} \int_{v \in S} #2 \dm{v}}

\begin{restatable}{lemma}{lemlowerbound}
\label{lem:lower-bound}
Any quantum algorithm for $\qsamplingab^\pp$ with $\pp \in [p_{\min}, p_{\max}]$ requires at least $\Omega(1/\norm{\waterfilling}_2)$ queries to $O$ and $O\ct$.
%
\end{restatable}

\begin{proof}
The proof proceeds as follows: we first define a subset of oracles that are sufficiently hard to distinguish to characterize the query complexity of the problem. Exploiting the automorphism group of this subset of oracles, we then show that any algorithm may be symmetrized in such a way that the real part of the amplitudes of the final state prepared by the algorithm does not depend on the specific oracle it was given. These amplitudes define a vector $\bar{\vc{\gamma}}$ that should satisfy some constraints for the algorithm to have success probability $p$. Moreover, we can use the hybrid argument to show that the components of $\bar{\vc{\gamma}}$ should also satisfy some constraints for the algorithm to be able to generate the corresponding state in at most $T$ queries. Putting all these constraints together in an optimization program, we then show that such a vector $\bar{\vc{\gamma}}$ cannot exist unless $T$ is large enough. This optimization program is then shown to be equivalent to the semidefinite program in Lemma~\ref{lem:sdp-solution}, which proves the theorem.

Let us now give the details of the proof. For given $\vc{\quantump}, \vc{\quantums} \in \R^n$, let us choose a subset of oracles $\O'_{\vc{\quantump},\vc{\quantums}}\subset\O_{\vc{\quantump},\vc{\quantums}}$ that are hard to distinguish. We choose the states hidden inside oracles to be of the form $\ket{\xi_k} = (-1)^{x_k} \0{d}$, where phases are given by some unknown string $\bv{x} \in \bb{n}$. We also choose $u$ so that any oracle in the subset acts trivially on the $d$-dimensional register holding the unknown states. In that case, this register is effectively one-dimensional, so we will omit it and write $(-1)^{x_k}$ as a relative phase.
More precisely, we consider a set of oracles $\O'_{\vc{\quantump},\vc{\quantums}} := \set{O_{\bv{x},u} : \bv{x} \in \bb{n}, u \in S}$, where
\begin{equation}
  S := \set{u \in \mathrm{U}(n) : u \0{n} = \0{n}} \cong \U{n-1}.
  \label{eq:S}
\end{equation}
As in the general case, we define the first oracle $O_{\bv{0},I}$ as a unitary that acts on $\0{n}$ as $O_{\bv{0},I} \0{n} = \ket{\statep}$, and arbitrarily on the orthogonal complement, and we use $O_{\bv{0},I}$ as a reference point to define the remaining oracles:
\begin{align*}
  O_{\bv{x},u} &:= V_{\bv{x}} \, O_{\bv{0},I} u, &
  \text{where} &&
  V_{\bv{x}} &:= \textstyle \sum_{k=1}^n\ (-1)^{x_k} \ket{k} \bra{k}.
\end{align*}
The set of target states is $\Psi'_{\vc{\quantump},\vc{\quantums}} := \set{\ket{\states(\bv{x})} : \bv{x} \in \bb{n}, u \in S}$ where $\ket{\states(\bv{x})} := V_{\bv{x}} \ket{\states} = \sum_{k=1}^n (-1)^{x_k} \quantums_k \ket{k}$. For the quantum state generation problem  corresponding to the restricted set of oracles $\O'_{\vc{\quantump},\vc{\quantums}}$, the automorphism group is $G = \bb{n} \times \U{n-1}$ and it acts on itself, \ie, $\X = G$. Note that the target states depend only on $\bv{x}$, but $u$ is used for parameterizing the oracles. Intuitively, the reason we need this parameter is that the algorithm should not depend on how the black box acts on states other than $\0{n}$ (or how its inverse acts on states other than $\ket{\statepxi}$). To make this intuition formal, we will later choose the parameter $u$ for different oracles adversarially.

Let us consider an algorithm that uses $T$ queries to the black box $O_{\bv{x},u}$ and its inverse, and let us denote the final state of this algorithm by $\ket{\psi_T(\bv{x},u)}$. If we expand the first register in the standard basis, we can express this state as
\begin{equation*}
  \ket{\psi_T(\bv{x},u)} = \textstyle \sum_{k=1}^n (-1)^{x_k} \ket{k} \ket{\g{k}{\bv{x}}{u}}.
\end{equation*}
Here the workspace vectors $\ket{\g{k}{\bv{x}}{u}}$ can have arbitrary dimension and are not necessarily unit vectors, but instead satisfy the normalization constraint $\sum_{k=1}^n \|\ket{\g{k}{\bv{x}}{u}}\|_2^2 = 1$. If the algorithm succeeds with probability $\pp$, then according to Eq.~(\ref{eq:Closeness}) for any $\bv{x}$ and $u$ we have
\begin{align*}
  \sqrt{\pp}
  &\leq   \Re \bigl[\bra{\states(\bv{x})} \bra{\bar{0}} \cdot \ket{\psi_T(\bv{x},u)}\bigr]\\
  &= \Re \bigl[\textstyle \sum_{k=1}^n \quantums_k\cdot
     \braket{\bar{0}}{\g{k}{\bv{x}}{u}}\bigr]\\
  &= \vc{\quantums}\tp \cdot \vc{\gamma}(\bv{x},u),
\end{align*}
where $\vc{\gamma}(\bv{x},u)$ is a real vector whose components are given by
\begin{equation*}
  \g{k}{\bv{x}}{u} := \Re \bigl[ \braket{\bar{0}}{\g{k}{\bv{x}}{u}} \bigr ].
\end{equation*}
Note that $\norm{\vc{\gamma}(\bv{x},u)}_2 \leq 1$.

\begin{figure*}
  
  \centering
  \TikZorPDF{
    \input{fig-symmetrization.tex}
  }{
    \includegraphics[width=0.95\textwidth]{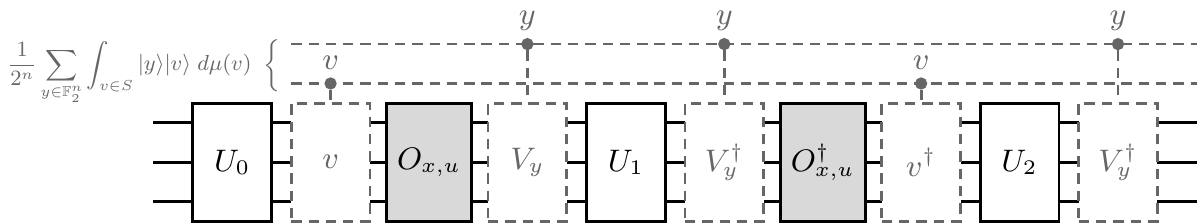}
  }

  \caption{Symmetrized algorithm. We symmetrize the algorithm by introducing unitaries $V_{\bv{y}}$ and $v$, controlled by an extra register prepared in the uniform superposition over all $\ket{\bv{y}}$ and $\ket{v}$.}
  \label{fig:Symmetrization}
\end{figure*}

Next, let us show that we can symmetrize the algorithm without decreasing its success probability. We do this by replacing each oracle call by $O_{\bv{x}+\bv{y},uv} = V_{\bv{y}} O_{\bv{x},u} v$ and correcting the final state by applying $V_{\bv{y}}\ct$ (see Fig.~\ref{fig:Symmetrization}). Let $\mu$ be the Haar measure on the set $S$ defined in Eq.~(\ref{eq:S}). We define an operation that symmetrizes a set of states:
\begin{equation*}
  \overline{\ket{\phi(\bv{x},u)}}\!
  := \!\!\twirling{
       \!\Bigl[ (V_{\bv{y}}\ct \x I) \ket{\phi(\bv{x} + \bv{y},uv)} \Bigr]
       \ket{\bv{y}} \ket{v}\!
     }.
\end{equation*}
If we symmetrize the final state $\ket{\psi_T(\bv{x},u)}$, we get
\begin{multline*}
  \overline{\ket{\psi_T(\bv{x},u)}} = \\
  \twirling{
    \sum_{k=1}^n (-1)^{x_k}
      \ket{k} \ket{\g{k}{\bv{x} + \bv{y}}{uv}}
      \ket{\bv{y}} \ket{v}
  }.
\end{multline*}
Note that the target state $\ket{\states(\bv{x})} \0{}$ is already symmetric, so symmetrization only introduces an additional workspace register in a default state (uniform superposition):
\begin{equation*}
  \overline{\ket{\states(\bv{x})} \0{}}
   =  \ket{\states(\bv{x})}\0{}\twirling{ \ket{\bv{y}} \ket{v}
      }.
\end{equation*}
The success probability of the symmetrized algorithm is
\begin{align*}
  \sqrt{\bar{\pp}}
  &:=   \Re \Bigl[\overline{\bra{\states(\bv{x})} \bra{\bar{0}}}
            \cdot \overline{\ket{\psi_T(\bv{x},u)}} \Bigr] \\
  & = \sum_{k=1}^n \quantums_k \cdot
      \twirlingsquare{\Re \bigl[\braket{\bar{0}}{\g{k}{\bv{x} + \bv{y}}{uv}}}\bigr] \\
  & = \vc{\quantums}\tp \cdot \bar{\vc{\gamma}},
\end{align*}
where, by changing variables, we get that $\bar{\vc{\gamma}}$ is the average of vectors $\vc{\gamma}(\bv{y},v)$ and thus does not depend on $\bv{x}$ and $u$:
\begin{equation*}
  \bar{\vc{\gamma}} := \twirlingsquare{\vc{\gamma}(\bv{y},v)}.
\end{equation*}
Note that $\norm{\bar{\vc{\gamma}}}_2 \leq 1$ by triangle inequality. Also, note that $\bar{\pp} \geq \pp$, since the mean is at least as large as the minimum. Thus without loss of generality we can consider only symmetrized algorithms.

Let $\bv{x}, \bv{x}' \in \bb{n}$ and $u, u' \in S$. The difference of final states of the symmetrized algorithm that uses oracles $O_{\bv{x},u}$ and $O_{\bv{x}',u'}$ is given in Eq.~(\ref{eq:Distance lower bound}) on the next page.
\begin{figure*}
\normalsize
\setcounter{MYtempeqncnt}{\value{equation}}
\setcounter{equation}{9}
\begin{align}
    \norm{\overline{\ket{\psi_T(\bv{x},u)}} - \overline{\ket{\psi_T(\bv{x}',u')}}}_2^2
 &= \norm{
      \sum_{k=1}^n
      \twirling{
        \ket{k}
        \bigl(
          (-1)^{x _k} \ket{\g{k}{\bv{x} + \bv{y}}{u v}} -
          (-1)^{x'_k} \ket{\g{k}{\bv{x}'+ \bv{y}}{u'v}}
        \bigr)
        \ket{\bv{y}} \ket{v}
      }
    }_2^2 \nonumber \\
 &= \sum_{k=1}^n
    \twirlingsquare{
      \norm{
        (-1)^{x _k} \ket{\g{k}{\bv{x} + \bv{y}}{u v}} -
        (-1)^{x'_k} \ket{\g{k}{\bv{x}'+ \bv{y}}{u'v}}
      }_2^2
    } \nonumber \\
 &\geq
    \sum_{k=1}^n
    \twirlingsquare{
      \bigl(
        (-1)^{x _k} \g{k}{\bv{x} + \bv{y}}{u v} -
        (-1)^{x'_k} \g{k}{\bv{x}'+ \bv{y}}{u'v}
      \bigr)^2
    } \nonumber \\
 &\geq
    \sum_{k=1}^n
    \bigl(
      (-1)^{x _k} \gb{k} -
      (-1)^{x'_k} \gb{k}
    \bigr)^2
  = \sum_{k: x_k \neq x'_k} \bigl( 2 \gb{k} \bigr)^2.
  \label{eq:Distance lower bound}
\end{align}
\setcounter{equation}{\value{MYtempeqncnt}}
Here the two inequalities were obtained from the following facts:
\begin{enumerate}
\item If $\0{}$ is a unit vector then for any $\ket{\gamma}$ we have:
$
  \norm{\ket{\gamma}}_2^2 \geq
  \norm{\0{} \braket{\bar{0}}{\gamma}}_2^2 =
  \abs {\braket{\bar{0}}{\gamma}}^2 \geq
  \bigl(\Re[\braket{\bar{0}}{\gamma}]\bigl)^2.
$
\item For any function $\gamma(\bv{y},v)$ by Cauchy--Schwarz inequality we have:
\begin{equation*}
  \twirlingsquare{\gamma(\bv{y},v)^2}
  \geq \Biggl( \twirlingsquare{\gamma(\bv{y},v)} \Biggr)^2.
\end{equation*}
\end{enumerate}
\hrulefill
\vspace*{4pt}
\end{figure*}%
\newcommand{\hybrid}{It follows by induction from the following fact. If $O$ and $O'$ are unitary matrices, then for any vectors $\ket{\psi}$ and $\ket{\psi'}$ it holds that
\begin{align*}
  \norm{O \ket{\psi} - O' \ket{\psi'}}_2
  &=    \norm{O \ket{\psi} - O' \ket{\psi} + O' \ket{\psi} - O' \ket{\psi'}}_2 \\
  &\leq \norm{(O - O') \ket{\psi}}_2 + \norm{O' (\ket{\psi} - \ket{\psi'})}_2 \\
  &\leq \norm{O - O'}_\infty + \norm{\ket{\psi} - \ket{\psi'}}_2.
\end{align*}
}%
By the hybrid argument~\cite{bbbv97,Vazirani:98}, we get the following upper bound:\footnote{\hybrid}
\begin{equation}
  \norm{\overline{\ket{\psi_T(\bv{x},u)}} - \overline{\ket{\psi_T(\bv{x}',u')}}}_2
  \leq T \cdot \norm{O_{\bv{x},u}-O_{\bv{x}',u'}}_\infty,
  \label{eq:Distance upper bound}
\end{equation}
where $\norm{\cdot}_\infty$ denotes the usual operator norm.
Bounds~(\ref{eq:Distance lower bound}) and~(\ref{eq:Distance upper bound}) together imply that for any $\bv{x}, \bv{x}' \in \bb{n}$ and $u, u' \in S$:
\begin{equation}
  T \geq \frac{\sqrt{\sum_{k: x_k \neq x'_k} \bigl( 2 \gb{k} \bigr)^2}}
                {\norm{O_{\bv{x},u}-O_{\bv{x}',u'}}_\infty}.
  \label{eq:Lower bound}
\end{equation}
To obtain a good lower bound, we want to choose oracles $O_{\bv{x},u}$ and $O_{\bv{x}',u'}$ to be as similar as possible. First, let us fix $u := I$, $\bv{x} := \bv{0}$ and $\bv{x}' := \bv{e}_l$, where $\bv{e}_l$ is the $l$-th standard basis vector. Then, the numerator in Eq.~(\ref{eq:Lower bound}) is simply $2\gb{l}$. Let us choose $u'$ in order to minimize the denominator. Recall that $O_{\bv{0},I} \0{n} = \ket{\statep}$ and note that any unitary matrix that fixes $\ket{\statep}$ can be written as $O_{\bv{0},I} u' (O_{\bv{0},I})\ct$ for some choice of $u'$ fixing $\0{n}$. Since $O_{\bv{x}',u'} \0{n} = V_{\bv{e}_l} O_{\bv{0},I} u' \0{n} = V_{\bv{e}_l} \ket{\statep}$, we also have $O_{\bv{x}',u'} (O_{\bv{0},I})\ct \ket{\statep} = V_{\bv{e}_l} \ket{\statep}$, and any unitary matrix that sends $\ket{\statep}$ to $V_{\bv{e}_l} \ket{\statep}$ can be expressed as $O_{\bv{x}',u'} (O_{\bv{0},I})\ct$ for some choice of $u'$. Let us choose $u'$ so that $O_{\bv{x}',u'} (O_{\bv{0},I})\ct$ acts as a rotation in the two-dimensional subspace $\spn \set{\ket{\statep}, V_{\bv{e}_l} \ket{\statep}}$ and as identity on the orthogonal complement. If $\theta$ denotes the angle of this rotation, then $\cos \theta = \bra{\vc{\quantump}} V_{\bv{e}_l} \ket{\statep} = \sum_{k=1}^n \quantump_k^2 - 2 \quantump_l^2 = 1 - 2 \quantump_l^2$ and $\sin \theta = \sqrt{1 - (1 - 2 \quantump_l^2)^2} = 2 \quantump_l \sqrt{1- \quantump_l^2}$. Then
\begin{align*}
\norm{O_{\bv{0},I} - O_{\bv{x}',u'}}_\infty
&= \norm{I - O_{\bv{x}',u'} (O_{\bv{0},I})\ct}_\infty \\
&= \norm{I -
    \mx{
      \cos \theta & -\sin \theta \\
      \sin \theta &  \cos \theta
    }
  }_{\infty}\\
&= 2 \quantump_l
  \norm{
    \mx{
      \quantump_l & \sqrt{1 - \quantump_l^2} \\
      -\sqrt{1 - \quantump_l^2} &  \quantump_l
    }
  }_{\infty} \\
&= 2 \quantump_l,
\end{align*}
where the singular values of the last matrix are equal to $1$, since it is a rotation. By plugging this back in Eq.~(\ref{eq:Lower bound}), we get that for any $l \in [n]$:
\begin{equation*}
  T \geq \frac{\abs{\gb{l}}}{\quantump_l}.
\end{equation*}

Thus any quantum algorithm that solves $\qsamplingab^\pp$ with $T$ queries and success probability $\pp$ gives us some vector $\bar{\vc{\gamma}}$ such that
\begin{align}
  \norm{\bar{\vc{\gamma}}}_2 &\leq 1, &
  \vc{\quantums}\tp \cdot \bar{\vc{\gamma}} &\geq \sqrt{\pp}, &
  \forall l: \abs{\gb{l}} \leq T \quantump_l.
  \label{eq:Feasibility}
\end{align}
To obtain a lower bound on $T$, we have to find the smallest possible $t$ such that there is still a feasible value of $\bar{\vc{\gamma}}$ that satisfies Eqs.~(\ref{eq:Feasibility}) (with $T$ replaced by $t$).
We can state this as an optimization problem:
\begin{equation}
\begin{array}{rl}
  T \geq \min_{\bar{\vc{\gamma}}} t
  \quad \text{s.t.} \quad
  & \norm{\bar{\vc{\gamma}}}_2 \leq 1, \\
  & \forall l: \abs{\gb{l}} \leq t \quantump_l, \\
  & \vc{\quantums}\tp \cdot \bar{\vc{\gamma}} \geq \sqrt{\pp}.
\end{array}
\label{eq:Subnormal gamma}
\end{equation}

Finally, let us show that we can start with a feasible solution $\bar{\vc{\gamma}}$ of problem~(\ref{eq:Subnormal gamma}) and modify its components, without increasing the objective value or violating any of the constraints, so that $\forall l: \gb{l} \geq 0$ and $\norm{\bar{\vc{\gamma}}}_2 = 1$. Clearly, making all components of $\bar{\vc{\gamma}}$ non-negative does not affect the objective value and makes the last constraint only easier to satisfy since $\quantums_k\geq 0$ for all $k$. To turn $\bar{\vc{\gamma}}$ into a unit vector, first observe that not all of the constraints $\gb{l} \leq t \quantump_l$ can be saturated (indeed, in that case we would have $\bar{\vc{\gamma}} = t \vc{\quantump}$ with $t < 1$ since $\norm{\bar{\vc{\gamma}}}_2 < \norm{\vc{\quantump}}_2 = 1$, but the last constraint then implies $\vc{\quantums}\tp \cdot \vc{\quantump} > \sqrt{\pp}$, which contradicts the assumption $\pp \geq p_{\min}$). If $\norm{\bar{\vc{\gamma}}}_2 < 1$, let $j$ be such that $\gb{j} < t \quantump_j$. We increase $\gb{j}$ until either $\norm{\bar{\vc{\gamma}}}_2 = 1$ or $\gb{j} = t \quantump_j$. We then repeat with another $j$ such that $\gb{j} < t \quantump_j$, until we reach $\norm{\bar{\vc{\gamma}}}_2 = 1$. Note that while doing so, the other constraints remain satisfied. Therefore, the program reduces to
\begin{equation*}
\begin{array}{rl}
  T \geq \min_{\bar{\vc{\gamma}}} t
  \quad \text{s.t.} \quad
  & \norm{\bar{\vc{\gamma}}}_2 = 1, \\
  & \forall l: 0 \leq \gb{l} \leq t \quantump_l, \\
  & \vc{\quantums}\tp \cdot \bar{\vc{\gamma}} \geq \sqrt{\pp}.
\end{array}
\end{equation*}
Note that we need $p\leq p_{\max}$, otherwise this program has no feasible point. Setting $\ve = \bar{\vc{\gamma}}/t$, we may rewrite this program as in Eq.~(\ref{eq:Primal1}) in Appendix~\ref{apx:SDP}:
\begin{equation*}
\begin{array}{rl}
  \frac{1}{T^2}\leq\max_{\e_k \geq 0} \norm{\ve}_2^2
  \quad \text{s.t.} \quad
  & \forall k: \quantump_k \geq \e_k \geq 0, \\
  & \vc{\quantums}\tp \cdot \ve \geq \sqrt{\pp} \norm{\ve}_2,
\end{array}
\end{equation*}
Finally, setting $M=\ve\cdot\ve\tp$, this program becomes the same as the SDP in Eq.~(\ref{eq:Primal}), with the additional constraint that $M$ is rank-1. However, we know from Lemma~\ref{lem:sdp-solution} that the SDP in Eq.~(\ref{eq:Primal}) admits a rank-1 optimal point, therefore adding this constraint does not modify the objective value, which is also $\norm{\waterfilling}_2^2$ by Lemma~\ref{lem:sdp-solution}.
\qed
\end{proof}

\section{Quantum rejection sampling\\algorithm} \label{sect:Algorithm}


In this section, we describe quantum rejection sampling algorithms for $\qsamplingab$ and $\sqsamplingab$ problems. We also explain the intuition behind our method and its relation to the classical rejection sampling. Our algorithms are based on amplitude amplification~\cite{AmplitudeAmplification} and can be seen as an extension of the algorithm in~\cite{Grover00}.

\subsection{Intuitive description of the algorithm} 

The workspace of our algorithm is $\C^d \x \C^n \x \C^2$, where the last register can be interpreted as a quantum coin that determines whether a sample will be rejected or accepted (this corresponds to basis states $\ket{0}$ and $\ket{1}$, respectively). Our algorithm is parametrized by a vector $\ve \in \R^n$ ($0 \leq \e_k \leq \quantump_k$ for all $k$) that characterizes how much of the amplitude from the initial state will be used for creating the final state (in classical rejection sampling $\e_k^2$ corresponds to the probability that a specific value of $k$ is drawn from the initial distribution and accepted).

We start in the initial state $\0{d}\0{n}\ket{0}$ and apply the oracle $O$ to prepare $\ket{\statepxi}$ in the first two registers:
\addtocounter{equation}{1}
\begin{equation}
  O \0{dn} \x \ket{0}
  = \ket{\statepxi} \ket{0}
  = \sum_{k=1}^n \quantump_k \ket{\xi_k} \ket{k} \ket{0}.
  \label{eq:a0}
\end{equation}
Next, for each $k$ let $\Rotk$ be a single-qubit unitary operation defined\footnote{For those $k$, for which $\quantump_k = 0$, operation $\Rotk$ can be defined arbitrarily.} as follows (this is a rotation by an angle whose sine is equal to $\e_k/\quantump_k$):
\begin{equation}
  \Rotk := \frac{1}{\quantump_k}
  \mx{
    \sqrt{\abs{\quantump_k}^2 - \e_k^2} &
   -\e_k \\
    \e_k &
    \sqrt{\abs{\quantump_k}^2 - \e_k^2}
  }.
  \label{eq:Rotk}
\end{equation}
Let $\Rot := \sum_{k=1}^n \ket{k} \bra{k} \x \Rotk$ be a block-diagonal matrix that performs rotations by different angles in mutually orthogonal subspaces. Then $I_d \x \Rot$ corresponds to applying $\Rotk$ on the last qubit, controlled on the value of the second register being equal to $k$. This operation transforms state~(\ref{eq:a0}) into
\begin{align}
  \ket{\Psi_{\ve}}
 &:= (I_d \x \Rot) \cdot \ket{\statepxi} \ket{0} \nonumber \\
 & = \sum_{k=1}^n \ket{\xi_k} \ket{k}
     \Bigl( \sqrt{\abs{\quantump_k}^2 - \e_k^2} \, \ket{0} + \e_k \ket{1} \Bigr).
  \label{eq:Psie}
\end{align}

If we would measure the coin register of state $\ket{\Psi_{\ve}}$ in the basis $\set{\ket{0}, \ket{1}}$, the probability of outcome $\ket{1}$ (``accept'') and the corresponding post-measurement state would be
\begin{align}
  \qe &:= \norm{\bigl(I_d \x I_n \x \ket{1} \bra{1}\bigr) \ket{\Psi_{\ve}}}_2^2
       = \sum_{k=1}^n \e_k^2
       = \norm{\ve}_2^2, \label{eq:qe} \\
  \ket{\Psi_{\Pi,\ve}}
      &:= \frac{1}{\norm{\ve}_2} \sum_{k=1}^n \e_k \ket{\xi_k} \ket{k} \ket{1}.\nonumber
\end{align}
Note that if the vector of amplitudes $\ve$ is chosen to be close to $\vc{\quantums}$, then the reduced state on the first two registers of $\ket{\Psi_{\Pi,\ve}}$ has a large overlap on the target state $\ket{\statesxi}$, more precisely,
\begin{equation}
  \sqrt{\pe} :=  \bigl(\bra{\statesxi} \x \bra{1} \bigr) \ket{\Psi_{\Pi,\ve}}
       = \vc{\quantums}\tp \cdot \frac{\ve}{\norm{\ve}_2},
  \label{eq:pe}
\end{equation}
Depending on the choice of $\ve$, this can be a reasonably good approximation, so our strategy will be to prepare a state close to $\ket{\Psi_{\Pi,\ve}}$.

In principle, we could obtain $\ket{\Psi_{\Pi,\ve}}$  by repeatedly preparing $\ket{\Psi_{\ve}}$ and measuring its coin register until we get the outcome ``accept'' (we would succeed with high probability after $O(1/\qe)$ steps). To speed up this process, we can use amplitude amplification~\cite{AmplitudeAmplification} to amplify the amplitude of the ``accept'' state $\ket{1}$ in the coin register of the state in Eq.~(\ref{eq:Psie}). This allows us to increase the probability of outcome ``accept'' arbitrarily close to $1$ in $O(1/\sqrt{\qe})$ steps.

\subsection{Amplitude amplification subroutine and\\quantum rejection sampling algorithm}

\begin{figure}[th]
  
  \centering
  \TikZorPDF{
    \input{fig-Ue.tex}
  }{
    \includegraphics[width=0.20\textwidth]{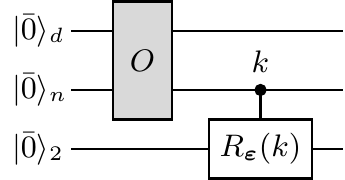}
  }

  \caption{Quantum circuit for implementing $\Ue$.}
  \label{fig:Ue}
\end{figure}

We will use the following amplitude amplification subroutine extensively in all algorithms presented in this paper:
\begin{equation*}
  \SQRS(\reflection{\ket{\statepxi}\ket{0}},\ve,t) :=
  \bigl(
    \reflection{\ket{\Psi_{\ve}}} \cdot
    \reflection{I_d \x I_n \x \ket{1} \bra{1}}
  \bigl)^t,
\end{equation*}
where reflections through the two subspaces are defined as follows:
\begin{align*}
  \reflection{I_d \x I_n \x \ket{1} \bra{1}}
   &:= I_d \x I_n \x \bigl( I_2 - 2 \ket{1} \bra{1} \bigr)
     = I_d \x I_n \x Z, \\
  \reflection{\ket{\Psi_{\ve}}}
   &:= (I_d \x \Rot) \; \reflection{\ket{\statepxi}\ket{0}} \; (I_d \x \Rot) \ct.
\end{align*}
Depending on the application, we will either be given an oracle $O$ for preparing $\ket{\statepxi} \ket{0}$ or an oracle $\reflection{\ket{\statepxi}\ket{0}}$ for reflecting through this state. Note that we can always use the preparation oracle to implement the reflection $\reflection{\ket{\statepxi}\ket{0}}$ as
\begin{equation}
  (O \x I_2) \;
  \bigl( I_d \x I_n \x I_2 - 2 \ket{\bar{0},\bar{0},0} \bra{\bar{0},\bar{0},0} \bigl) \;
  (O \x I_2)\ct.
  \label{eq:refpxi}
\end{equation}

\algorithm{
  \textbf{Amplitude amplification subroutine}
  $\SQRS(\reflection{\ket{\statepxi}\ket{0}}, \ve, t)$
  \textbf{for quantum rejection sampling} \\
  Perform the following steps $t$ times:
  \begin{enumerate}
    \item Perform $\reflection{I_d \x I_n \x \ket{1} \bra{1}}$
          by applying Pauli $Z$ on the coin register.
    \item Perform $\reflection{\ket{\Psi_{\ve}}}$ by applying $\Rot\ct$
          on the last two registers, applying $\reflection{\ket{\statepxi}\ket{0}}$,
          and then undoing $\Rot$.
  \end{enumerate}
}

The quantum rejection sampling algorithm $\AQRS(O,\statep,\ve)$ starts with the initial state $\0{d} \0{n} \ket{0}$. First, we transform it into $\ket{\Psi_{\ve}}$ defined in Eq.~(\ref{eq:Psie}), by applying $\Ue := (I_d \x \Rot) \cdot (O \x I_2)$ (see Fig.~\ref{fig:Ue}). Then we apply the amplitude amplification subroutine $\SQRS(\reflection{\ket{\statepxi}\ket{0}}, \ve, t)$ with $t = O\bigl(1/\norm{\ve}_2\bigr)$. Finally, we measure the last register: if the outcome is $\ket{1}$, we output the first two registers, otherwise we output ``Fail''. To prevent the outcome ``Fail'' we can slightly adjust the angle of rotation in amplitude amplification so that the target state is reached exactly. More precisely, we prove that one can choose $\ve = r \cdot \waterfilling$ for some bounded constant $r$, so that amplitude amplification succeeds with probability $1$ (\ie, the outcome of the final measurement is always $\ket{1}$). Moreover, such algorithm is optimal as its cost matches the lower bound in Lemma~\ref{lem:lower-bound}.

\algorithm{
  \textbf{Quantum rejection sampling algorithm}
  $\AQRS(O, \statep, \ve)$
  \begin{enumerate}
    \item Start in initial state $\0{d} \0{n} \ket{0}$.
    \item Apply $\Ue$.
    \item On the current state apply the amplitude amplification subroutine
          $\SQRS(\reflection{\ket{\statepxi}\ket{0}}, \ve, t)$
          where $\reflection{\ket{\statepxi}\ket{0}}$
          is implemented according to Eq.~(\ref{eq:refpxi}) and
          $t = O(1/\norm{\ve}_2)$.
    \item Measure the last register. If the outcome is $\ket{1}$,
          output the first two registers, otherwise output ``Fail''.
  \end{enumerate}
}

\begin{lemma}\label{lem:algo-sdp}
For any $p_{\min}\leq \pp\leq p_{\max}$, there is a constant $r \in [\frac{1}{2},1]$, so that the algorithm $\AQRS(O, \statep, \ve)$ with $\ve = r \cdot \waterfilling$ solves $\qsamplingab$ with success probability $\pp$ using $O \bigl( 1 / \norm{\waterfilling}_2 \bigr)$ queries to $O$ and $O\ct$.
\end{lemma}

\begin{proof}
By Def.~\ref{def:water-filling}, we have $0 \leq \e_k \leq \quantump_k$ for all $k$, therefore $\waterfilling$ is a valid choice of vector $\ve$ for the algorithm. Instead of using $\waterfilling$ itself, we slightly scale it down by a factor $r$ so that the amplitude amplification never fails. Note that if we were to use $\ve=\waterfilling$, the probability that the amplitude amplification succeeds after $t$ steps would be $\sin^2 \bigl( (2t+1) \theta \bigr)$, where $\theta := \arcsin\norm{\waterfilling}_2$ (see \eg~\cite{BoyerBHT98,AmplitudeAmplification} for details). Note that this probability would be equal to one at $t = \frac{\pi}{4\theta}-\frac{1}{2}$, which in general might not be an integer. However, following an idea from~\cite[p.10]{AmplitudeAmplification}, we can ensure this by slightly decreasing $\theta$ to $\tilde{\theta} := \frac{\pi}{2(2\tilde{t}+1)}$ where $\tilde{t} := \lceil\frac{\pi}{4\theta}-\frac{1}{2}\rceil$. This can be done by setting $\ve := r \cdot \waterfilling$ with the scaling-down factor $r := \frac{\sin\tilde{\theta}}{\sin\theta}$. One can check that $r$ satisfies $\frac{1}{2} \leq r \leq 1$ (this follows from $0 \leq \theta \leq \frac{\pi}{2}$).

Together with Eq.~(\ref{eq:pe}), Def.~\ref{def:water-filling} also implies that for this choice, the algorithm solves $\qsamplingab$ with success probability $\frac{\vc{\quantums}\tp \cdot \ve}{\norm{\ve}_2} = \frac{\vc{\quantums}\tp \cdot \waterfilling}{\norm{\waterfilling}_2} = \sqrt{\pp}$. It therefore remains to prove that the cost of the algorithm is $O(1/\norm{\ve}_2)$, which follows immediately from its description: we need one query to implement the operation $\Ue$ and two queries to implement $\reflection{\ket{\statepxi}\ket{0}}$, thus in total we need $2t+1 = O\bigl(1/\sqrt{\qe}\bigr) = O(1/\norm{\ve}_2)$ calls to oracles $O$ and $O\ct$.
\qed
\end{proof}

We now have all the elements to prove Theorem~\ref{thm:qsampling}.
\thmqsampling*

\begin{proof}
When $p\leq p_{\min}$, the state $\ket{\statepxi}$ is already close enough to $\ket{\statesxi}$ to satisfy the constraint on the success probability, therefore one call to $O$ is sufficient, which is clearly optimal. When $p> p_{\max}$, the oracle gives no information about some of the unknown states $\ket{\xi_k}$ (when $\quantump_k=0$), but the target state should have some overlap on $\ket{\xi_k}\ket{k}$ to satisfy the constraint on the success probability, therefore the problem is not solvable.

For the general case $p_{\min}\leq p\leq p_{\max}$, the upper bound comes from the algorithm in Lemma~\ref{lem:algo-sdp}, and the matching lower bound is given in Lemma~\ref{lem:lower-bound}.
\qed
\end{proof}

\subsection{Strong quantum rejection sampling algorithm}

Let us now describe how the algorithm can be modified to solve the stronger problem $\sqsamplingab$. The first modification follows from the observation that in the previous algorithm, the oracle is only used in two different ways: it is used once to create the state $\ket{\statepxi}$, and then only to reflect through that state. This means that we can still solve the problem if, instead of being given access to an oracle that maps $\0{dn}$ to $\ket{\statepxi}$, we are provided with one copy of $\ket{\statepxi}$ and an oracle that reflects through it.

In order to solve $\sqsamplingab$, we should also be able to deal with the case where we only know the ratios between the amplitudes $\quantump_k$ and $\quantums_k$ for each $k$, instead of the amplitudes themselves. As we will see, in that case we cannot use the algorithm given above anymore, as we do not know in advance how many steps of amplitude amplification are required. There are different approaches to solve this issue, one of them being to estimate $\qe$, and therefore the required number of steps, by performing a phase estimation on the amplitude amplification operator (this is sometimes referred to as amplitude estimation or quantum counting, see~\cite{BoyerBHT98,AmplitudeAmplification}). Another option, also proposed by~\cite{BoyerBHT98,AmplitudeAmplification}, is to repeat the algorithm successively with an increasing number of steps until it succeeds. One advantage of the first option would be that it provides an estimation of the initial acceptance probability $\qe$, which might be useful for some applications. Since this is not required for $\sqsamplingab$, we will rather describe an algorithm based on the second option, as it is more direct. Note that for both options, we need to adapt the algorithms in~\cite{BoyerBHT98,AmplitudeAmplification} as they require to use a fresh copy of the initial state after each failed attempt, whereas for $\sqsamplingab$ we only have one copy of that state. This issue can be solved by using the state left over from the previous unsuccessful measurement instead of a fresh copy of the state. More precisely, we can use the following algorithm.
\algorithm{
\textbf{Strong quantum rejection sampling algorithm}\\
$\ASQRS(\ket{\statepxi},\ve,c)$
\begin{enumerate}
  \item\label{item:sqrs-initialization}
    Append an extra qubit prepared in the state $\ket{0}$ to the
    input state $\ket{\statepxi}$, and apply $I_d \x \Rot$ on the resulting state.
  \item\label{item:sqrs-measurement-1}
    Measure the last register of the current state. If the outcome
    is $\ket{1}$, output the first two registers and stop. Otherwise,
    set $l=0$ and continue.
  \item\label{item:sqrs-new-attempt}
    Let $T_l:=\lceil c^l\rceil$ and pick an integer $t \in [T_l]$
    uniformly at random.
  \item\label{item:sqrs-qrs}
    On the current state apply the amplitude amplification subroutine
    $\SQRS(\reflection{\ket{\statepxi}\ket{0}}, \ve, t)$.
  \item\label{item:sqrs-measurement-2}
    Measure the last register of the current state. If the outcome
    is $\ket{1}$, output the first two registers and stop. Otherwise,
    increase $l$ by one and go back to Step~\ref{item:sqrs-new-attempt}.
\end{enumerate}
}

\begin{lemma}\label{lem:algo-sqsqmpling}
For any $\alpha \geq 1$, there is a quantum algorithm that solves $\sqsamplingab$ with success probability $p(\gamma)$ using an expected number of queries $O(1/\norm{\ve(\gamma)}_2)$, where $\gamma = \alpha \norm{\vc{\quantump} \circ \vc{\tau}}_2$. In particular, for $\alpha = 1$ the expected number of queries is $O(1/\norm{\vc{\quantump} \circ \vc{\tau}}_2)$ and the success probability is equal to one.
\end{lemma}
Here the parameter $\alpha$ allows us to control the trade-off between the success probability and the required number of queries. However, we cannot predict the actual values of both quantities, because they depend on $\vc{\quantump}$ and $\vc{\tau}$, but only $\vc{\tau}$ is known to us. The only exception is $\alpha=1$, when the success probability is equal to one. Also, increasing $\alpha$ above $1/(\min_{k:\tau_k > 0} \tau_k)$ will no longer affect the query complexity and success probability of the algorithm.

\begin{proof}
We show that for some choice of $c > 1$ and $\ve$ the algorithm $\ASQRS(\ket{\statepxi},\ve,c)$ described above solves the problem.

Let us first verify that we can actually perform all steps required in the algorithm. We need one copy of the state $\ket{\statepxi}$, which is indeed provided as an input for the $\sqsamplingab$ problem. Note that for applying $\Rot$ in Step~\ref{item:sqrs-initialization} it suffices to know only the ratio $\e_k/\quantump_k$ (see Eq.~(\ref{eq:Rotk})), which can be deduced from $\tau_k$ as follows. Let $\ve := r \cdot \ve(\gamma)$ for some $r < 1$, and recall from Def.~\ref{def:water-filling} and~\ref{def:SQSampling} that $\e_k(\gamma) = \min \set{\quantump_k, \gamma \quantums_k}$ and $\quantums_k = \quantump_k \tau_k / \norm{\vc{\quantump} \circ \vc{\tau}}_2$, respectively. Then
\begin{align}
  \frac{\e_k}{\quantump_k}
 &= r \min \set{1, \gamma \frac{\quantums_k}{\quantump_k}} \nonumber \\
 &= r \min \set{1, \gamma \frac{\tau_k}{\norm{\vc{\quantump} \circ \vc{\tau}}_2}} \nonumber \\
 &= r \min \set{1, \alpha \tau_k},
  \label{eq:r}
\end{align}
where we substituted $\gamma = \alpha \norm{\vc{\quantump} \circ \vc{\tau}}_2$ from the statement of the Lemma. Note that once $r$ is chosen, the final expression in Eq.~(\ref{eq:r}) is completely known. Finally, applying $\SQRS(\reflection{\ket{\statepxi}\ket{0}}, \ve, t)$ in Step~\ref{item:sqrs-qrs} also requires the ability to apply $\Rot$, as well as $\reflection{\ket{\statepxi}\ket{0}}$, which can be done by using one oracle query. Therefore, we have all we need to implement the algorithm.

We now show that the algorithm has success probability $p(\gamma)$. Recall from Eq.~(\ref{eq:Psie}) that Step~\ref{item:sqrs-initialization} of the algorithm prepares the state
\begin{align*}
  \ket{\Psi_{\ve}}
  &= \sum_{k=1}^n \ket{\xi_k} \ket{k}
    \Bigl( \sqrt{\abs{\quantump_k}^2 - \e_k^2} \, \ket{0} + \e_k \ket{1} \Bigr) \\
  &= \sin \theta \ket{\Psi_{\Pi,\ve}} + \cos \theta \ket{\Psi_{\Pi,\ve}^\perp},
\end{align*}
where $\theta := \arcsin{\norm{\ve}_2}$ and unit vectors
\begin{align*}
  \ket{\Psi_{\Pi,\ve}}
  &:= \frac{1}{\sin \theta}
      \sum_{k=1}^n \e_k \ket{\xi_k} \ket{k} \ket{1}, \\
  \ket{\Psi_{\Pi,\ve}^\perp}
  &:= \frac{1}{\cos \theta}
      \sum_{k=1}^n \sqrt{\abs{\quantump_k}^2 - \e_k^2} \ket{\xi_k} \ket{k} \ket{0}
\end{align*}
are orthogonal and span a $2$-dimensional subspace. In this subspace $\reflection{I_d \x I_n \x \ket{1} \bra{1}}$ and $\reflection{\ket{\Psi_{\Pi,\ve}}}$ act in the same way, so each iteration of the amplitude amplification subroutine consists of a product of two reflections that preserve this subspace, and $\SQRS(\reflection{\ket{\statepxi}\ket{0}}, \ve, t)$ corresponds to a rotation by angle $2 t \theta$. Measurements in Steps~\ref{item:sqrs-measurement-1} and~\ref{item:sqrs-measurement-2} either project on $\ket{\Psi_{\Pi,\ve}}$, when the outcome is $\ket{1}$, or on $\ket{\Psi_{\Pi,\ve}^\perp}$, when the outcome is $\ket{0}$. Therefore, the algorithm always outputs the first two registers of the state $\ket{\Psi_{\Pi,\ve}}$, and by Eq.~(\ref{eq:pe}), the success probability is $\pe = p(\gamma)$, as claimed. In particular, for $\alpha = 1$ from Eq.~(\ref{eq:r}) we get $\e_k = r \pi_k \tau_k$ as $\tau_k \leq 1$. Thus, $\ve = r (\vc{\quantump} \circ \vc{\tau})$ and since $\vc{\quantums} = (\vc{\quantump} \circ \vc{\tau}) / \norm{\vc{\quantump} \circ \vc{\tau}}_2$, we get $\pe = (\vc{\quantums}\tp \cdot \ve/\norm{\ve}_2)^2 = 1$.

Let us now bound the expected number of oracle queries. We follow the proof of Theorem 3 in~\cite{AmplitudeAmplification}, but there is an important difference: a direct analogue of the algorithm in~\cite[Theorem 3]{AmplitudeAmplification} would use a fresh copy of $\ket{\statepxi}$ each time the measurement fails to give a successful outcome, whereas in this algorithm we start from the state left over from the previous measurement, since we only have one copy of $\ket{\statepxi}$. Note that $\SQRS(\reflection{\ket{\statepxi}\ket{0}}, \ve, t)$ in Step~\ref{item:sqrs-qrs} is always applied on $\ket{\Psi_{\Pi,\ve}^\perp}$, since it is the post-measurement state corresponding to the unsuccessful outcome. Therefore, the state created by Step~\ref{item:sqrs-qrs} is $\sin (2t\theta) \ket{\Psi_{\Pi,\ve}} + \cos(2t\theta) \ket{\Psi_{\Pi,\ve}^\perp}$, and the next measurement will succeed with probability $\sin^2(2t\theta)$. Since $t$ is picked uniformly at random between $1$ and $T_l$, the probability that the $l$-th measurement fails is
\begin{align}
  p_l &= \frac{1}{T_l} \sum_{t=1}^{T_l} \cos^2 (2 t \theta) \nonumber \\
      &= \frac{1}{2} + \frac{1}{2 T_l} \sum_{t=1}^{T_l} \cos (4 t \theta) \nonumber \\
   &\leq \frac{1}{2} + \frac{1}{2 T_l \norm{\ve}_2},
   \label{eq:pl}
\end{align}
where the upper bound is obtained as follows:
\begin{align*}
      \sum_{t=1}^T \cos (4t\theta)
   &= \Re \biggl( e^{i 4 \theta} \sum_{t=0}^{T - 1} e^{i 4 t \theta} \biggr) \\
   &= \Re \biggl( e^{i 4 \theta} \cdot \frac{1 - e^{i 4 T \theta}}{1 - e^{i 4 \theta}} \biggr) \\
   &= \Re \biggl( e^{i 2(T+1)\theta} \cdot
      \frac{e^{-i 2 T \theta} - e^{i 2 T \theta}}
           {e^{-i 2   \theta} - e^{i 2   \theta}} \biggr) \\
   &= \cos \bigl( 2(T+1)\theta \bigr)
      \frac{\sin(2T\theta)}{\sin(2\theta)} \\
&\leq \frac{1}{\sin(2\theta)}
 \leq \frac{1}{\sin \theta}
    = \frac{1}{\norm{\ve}_2},
\end{align*}
where we forced the last inequality by picking $r := \sqrt{3}/2$, so that $\sin \theta = \norm{\ve}_2 \leq \sqrt{3}/2$ and thus $0 \leq \theta \leq \pi/3$. Recall from the algorithm that $T_l = \lceil c^l \rceil$ for some $c > 1$, so it is increasing and goes to infinity as $l$ increases. Let $\bar{T} := 1/(2 \Delta \norm{\ve}_2)$ for some $\Delta > 0$ and let $\bar{l}$ be the smallest integer such that $T_l \geq \bar{T}$ for all $l \geq \bar{l}$. Then according to Eq.~(\ref{eq:pl}) we get that $p_l \leq 1/2 + 1/(2 \bar{T} \norm{\ve}_2) = 1/2 + \Delta =: \bar{p}$ for all $l \geq \bar{l}$. Note that the $l$-th execution of the subroutine uses at most $2T_l$ oracle queries, so the expected number of oracle calls is at most $2T_0 + p_0 (2T_1 + p_1 (2T_2 + \ldots))$. This can be upper bounded by
\begin{align}
      \sum_{l=0}^{\bar{l}} 2T_l + \sum_{d=1}^{\infty} 2T_{\bar{l}+d} \bar{p}^d
   &= \sum_{l=0}^{\bar{l}} 2 \lceil c^l \rceil
    + \sum_{d=1}^{\infty}  2 \lceil c^{\bar{l}+d} \rceil \bar{p}^d \nonumber \\
&\leq 4 \Biggl(
        \sum_{l=0}^{\bar{l}} c^l
      + c^{\bar{l}} \sum_{d=1}^{\infty} (c \bar{p})^d
      \Biggr) \label{eq:expqueries} \\
  & = 4 \Biggl(
        \frac{c^{\bar{l}+1} - 1}{c - 1}
      + c^{\bar{l}} \frac{c \bar{p}}{1 - c \bar{p}}
      \Biggr) \nonumber \\
&\leq 4 c^{\bar{l}+1}
      \Biggl(
        \frac{1}{c - 1}
      + \frac{\bar{p}}{1 - c \bar{p}}
      \Biggr) \nonumber \\
&\leq \frac{2 c^2}{\Delta \norm{\ve}_2}
      \Biggl(
        \frac{1}{c - 1}
      + \frac{\bar{p}}{1 - c \bar{p}}
      \Biggr), \label{eq:expqueriesbound}
\end{align}
where the first and last inequality is obtained from the following two observations, respectively:
\begin{enumerate}
  \item $\lceil c^l \rceil = c^l + \delta$ for some $0 \leq \delta < 1$, so $T_l = \lceil c^l \rceil < c^l + 1 < 2c^l$ as $c > 1$.
  \item $c^{\bar{l}+1} \leq c^2 \lceil c^{\bar{l}-1} \rceil = c^2 T_{\bar{l}-1} < c^2 \bar{T} = c^2 / (2 \Delta \norm{\ve}_2)$ by the choice of $\bar{l}$.
\end{enumerate}
Finally, we have to make a choice of $c > 1$ and $\Delta > 0$, so that the geometric series in Eq.~(\ref{eq:expqueries}) converges, \ie, $c \bar{p} < 1$ or equivalently $c < 2/(1+2\Delta)$. By choosing $c := 8/7$ and $\Delta := 1/4$ we minimize the upper bound in Eq.~(\ref{eq:expqueriesbound}) and obtain $128/\norm{\ve}_2 = O(1/\norm{\ve(\gamma)}_2)$. In particular, for $\alpha = 1$ this becomes $O(1/\norm{\vc{\quantump} \circ \vc{\tau}}_2)$.
\qed
\end{proof}

\section{Applications} \label{sect:applications}

\subsection{Linear systems of equations} \label{sect:linear}

As a first example of application, we show that quantum rejection sampling was implicitly used in the quantum algorithm for linear systems of equations proposed by Harrow, Hassidim, and Lloyd~\cite{HHL:09}. This algorithm solves the following quantum state generation problem: given the classical description of an invertible $d \times d$ matrix $A$ and a unit vector $\ket{b} \in \C^d$, prepare the quantum state $\ket{x}/\norm{\ket{x}}_2$, where $\ket{x}$ is the solution of the linear system of equations $A \ket{x} = \ket{b}$. As shown in~\cite{HHL:09}, we can assume without loss of generality that $A$ is Hermitian. Similarly to classical matrix inversion algorithms, an important factor of the performance of the algorithm is the condition number $\kappa$ of $A$, which is the ratio between the largest and smallest eigenvalue of $A$. We will assume that all eigenvalues of $A$ are between $\kappa^{-1}$ and $1$, and we denote by $\ket{\psi_j}$ and $\lambda_j$ the eigenvectors and eigenvalues of $A$, respectively. We also define\footnote{We choose the global phase of each eigenvector $\ket{\psi_j}$ so that $b_j$ is real and non-negative.} the amplitudes $b_j := \braket{\psi_j}{b}$, so that $\ket{b} = \sum_{j=1}^d b_j \ket{\psi_j}$. Then, the problem is to prepare the state $\ket{x} := A^{-1}\ket{b} = \sum_{j=1}^d b_j \lambda_j^{-1} \ket{\psi_j}$ (up to normalization).

We now show how this problem reduces to the quantum state conversion problem $\sqsamplingab$. Since $A$ is Hermitian, we can use Hamiltonian simulation techniques~\cite{Berry2006,Childs2009a,Childs2011} to simulate the unitary operator $e^{iAt}$ on any state. Using quantum phase estimation~\cite{Kitaev1995,CEMM98} on the operator $e^{iAt}$, we can implement an operator $E_A$ that acts in the eigenbasis of $A$ as $E_A: \ket{\psi_j} \0{} \mapsto \ket{\psi_j} \ket{\lambda_j}$, where $\ket{\lambda_j}$ is a quantum state encoding an approximation of the eigenvalue $\lambda_j$. Here, we will assume that this can be done exactly, that is, we assume that $\ket{\lambda_j}$ is a computational basis state that exactly encodes $\lambda_j$ (we refer the reader to~\cite{HHL:09} for a detailed analysis of the error introduced by this approximation). Under this assumption, the problem reduces to a quantum state conversion problem that we will call $\qle$. Its definition requires fixing a set of possible eigenvalues $\Lambda_{\kappa} \subset [\kappa^{-1},1]$ of finite cardinality $n := \abs{\Lambda_{\kappa}}$. Let us denote the set of $d \times d$ Hermitian matrices by $\Herm(d)$ and the eigenvalues of $A$ by $\spec(A)$.

\begin{definition}\deftitle{Quantum linear system of equations}
$\qle$, the \emph{quantum linear system of equations problem} is a quantum state conversion problem $(\O, \Phi, \Psi, \X)$ with $\X := \set{(\ket{b},A) \in \C^d \times \Herm(d): \spec(A) \in \Lambda_{\kappa}^d}$, oracles in $\O$ being pairs $(O_{\ket{b},A}, E_A)$ where $O_{\ket{b},A} := \reflection{\ket{b}}$ and $E_A$ acts as $E_A: \ket{\psi_j} \0{n} \mapsto \ket{\psi_j} \ket{\lambda_j}$ where $\ket{\psi_j}$ are the eigenvectors of $A$, and the corresponding initial and target states being $\ket{b}$ and $A^{-1}\ket{b} / \norm{A^{-1}\ket{b}}_2$.
\label{def:qlse}
\end{definition}

Using Lemma~\ref{lem:algo-sqsqmpling} we can prove the following result.
\thmlinearequations*
\begin{proof}
Following~\cite{HHL:09}, the algorithm for this problem consists of three steps:
\begin{enumerate}
  \item Apply the phase estimation operation $E_A$ on $\ket{b}$
    to obtain the state $\sum_{j=1}^d b_j \ket{\psi_j} \ket{\lambda_j}$.
  \item\label{step:lse-qrs} Convert this state to
    $\sum_{j=1}^d w_j \ket{\psi_j} \ket{\lambda_j} / \norm{\vc{w}}_2$.
  \item Undo the phase estimation operation $E_A$ to obtain the target state
    $\sum_{j=1}^d w_j \ket{\psi_j} / \norm{\vc{w}}_2$.
\end{enumerate}
We see that Step~\ref{step:lse-qrs} is an instance of $\sqsamplingab$, where the basis states $\set{\ket{\lambda} : \lambda \in \Lambda_{\kappa}}$ of the phase estimation register play the role of the states $\ket{k}$ and the vector $\vc{\tau}$ of the ratios between the initial and final amplitudes is given by $\tau_{\lambda} := (\kappa \lambda)^{-1}$ (here the normalization factor $\kappa$ is to make sure that $\max_{\lambda} \tau_{\lambda} = 1$). The rest of the reduction is summarized in Table~\ref{table:reduction-lse}. Therefore, we can use the algorithm from Lemma~\ref{lem:algo-sqsqmpling} to perform Step~\ref{step:lse-qrs}.

\begin{table}
\begin{center}
\begin{tabular}{|c|c|}
\hline
$\sqsamplingab$ & $\qle$ \\
\hline
$\ket{k}$ & $\ket{\lambda}$ \\
$\pi_k$ & $\pi_\lambda :=
\begin{cases}
  b_j \textrm{ if } \lambda = \lambda_j \in \spec(A) \\
  0   \textrm{ if } \lambda \notin \spec(A)
\end{cases}$ \\
$\ket{\xi_k}$ & $\ket{\xi_\lambda} :=
\begin{cases}
  \ket{\psi_j} \textrm{ if } \lambda = \lambda_j \in \spec(A) \\
  \textrm{n/a if }           \lambda \notin \spec(A)
\end{cases}$ \\
$\tau_k$ & $\tau_\lambda := (\kappa\lambda)^{-1}$ \\
\hline
\end{tabular}
\end{center}
\caption{Reduction from Step~\ref{step:lse-qrs} in the linear system of equations algorithm to the quantum resampling problem \textnormal{$\sqsamplingab$}.}
\label{table:reduction-lse}
\end{table}

If we set $\alpha := \kappa / \tilde{\kappa}$ then from Table~\ref{table:reduction-lse} we get
\begin{align*}
  \e_{\lambda_j}(\gamma)
 &= \quantump_{\lambda_j} \min \set{1, \alpha \tau_{\lambda_j}} \\
 &= b_j \min \set{1, (\tilde{\kappa} \lambda_j)^{-1}} \\
 &= \frac{b_j}{\tilde{\kappa} \max \set{\tilde{\kappa}^{-1}, \lambda_j}} \\
 &= \frac{\tilde{w}_j}{\tilde{\kappa}},
\end{align*}
thus $\ve(\gamma) = \tilde{\vc{w}} / \tilde{\kappa}$ and the expected number of queries is $O(1/\norm{\ve(\gamma)}_2) = O(\tilde{\kappa} / \norm{\tilde{\vc{w}}}_2)$. Recall that the amplitudes of the target state are given by $\quantums_j = w_j / \norm{\vc{w}}_2$, so $\quantums = \vc{w} / \norm{\vc{w}}_2$ and the success probability is
\begin{equation*}
  p = \biggl( \frac{\vc{\quantums}\tp \cdot \ve(\gamma)}{\norm{\ve(\gamma)}_2} \biggr)^2
    = \frac{\vc{w}\tp}{\norm{\vc{w}}_2} \cdot
      \frac{\tilde{\vc{w}}\;}{\norm{\tilde{\vc{w}}}_2}
\end{equation*}
as claimed.
\qed
\end{proof}

Note that even though we have a freedom to choose $\tilde{\kappa}$, we cannot predict the query complexity in advance, since it depends on $\tilde{w}_j = \braket{\psi_j}{b} / \tilde{\lambda}_j$, which in turn is determined by the lengths of projections of $\ket{b}$ in the eigenspaces of $A$, weighted by the corresponding truncated eigenvalues $\tilde{\lambda}_j$. Similarly, we cannot predict the success probability~$p$. However, by choosing $\tilde{\kappa} = \kappa$ we can at least make sure that $p = 1$ (since $\tilde{\lambda}_j = \lambda_j$ and thus $\tilde{\vc{w}} = \vc{w}$). In this case Step~\ref{step:lse-qrs} is performed exactly (assuming an ideal phase estimation black box) and the expected number of queries is $O(\kappa / \norm{\vc{w}}_2)$. By noting that $\lambda_j \leq 1$ for all $j$, we see that $\norm{\vc{w}}_2^2 = \sum_{j=1}^d b_j^2 \lambda_j^{-2} \geq 1$ and thus we can put a cruder upper bound of $O(\kappa)$, which coincides with the bound given in~\cite{HHL:09} for that step of the algorithm. For ill-conditioned matrices, \ie, matrices with a high condition number $\kappa$, the approach taken by~\cite{HHL:09} is to ignore small eigenvalues $\lambda_j\leq\tilde{\kappa}^{-1}$, for some cut-off $\tilde{\kappa}^{-1} \geq \kappa^{-1}$, which reduces the cost of the algorithm to $O(\tilde{\kappa})$, but introduces some extra error. In our case, by choosing $\alpha = \kappa / \tilde{\kappa}$ we obtained bound $O(\tilde{\kappa} / \norm{\tilde{\vc{w}}}_2)$, where $\norm{\tilde{\vc{w}}}_2 \geq 1$. Again, here $\tilde{\vc{w}}$ depends on additional structure of the problem and cannot be predicted beforehand.

In practical applications, we will of course not be given access to the ideal phase estimation operator $E_A$, but we can still approximate it by using the phase estimation algorithm~\cite{Kitaev1995,CEMM98} on the operator $A$. It is shown in~\cite{HHL:09} that if $A$ is $s$-sparse, this approximation can be implemented with sufficient accuracy at a cost $\tilde{O} \bigl( \log(d) s^2 \tilde{\kappa} / \epsilon \bigr)$, where $\epsilon$ is the overall additive error introduced by this approximation throughout the algorithm. Therefore, the total cost of the algorithm is at most $\tilde{O} \bigl( \log(d) s^2 \tilde{\kappa}^2 / \epsilon \bigr)$ (see~\cite{HHL:09} for details).

\subsection{Quantum Metropolis sampling} \label{sect:metropolis}

Since rejection sampling lies at the core of the (classical) Metropolis algorithm, it seems natural to use quantum rejection sampling to solve the corresponding problem in the quantum case. The quantum Metropolis sampling algorithm presented in~\cite{TOV+:2009} follows the same lines as the classical algorithm by setting up a (classical) random walk between eigenstates of the Hamiltonian, where each move is either accepted or rejected depending on the value of some random coin. The main complication compared to the classical version comes from the case where the move has to be rejected, since we cannot keep a copy of the previous eigenstate due to the no-cloning theorem. The solution proposed by Temme~\textit{et al.}~\cite{TOV+:2009} is to use an unwinding technique based on successive measurements to revert to the original state. Here, we show that quantum rejection sampling can be used to avoid this step, as it allows to amplify the amplitude of the ``accept'' state of the coin register, effectively eliminating rejected moves. This yields a more efficient algorithm as it eliminates the cost of reverting rejected moves and provides a quadratic speed-up on the overall cost of obtaining an accepted move.\footnote{Martin Schwarz has pointed out to us that this is similar to how~\cite{NWZ09} provides a speed-up over~\cite{MW05}, and that our technique can also be used to speed-up the quantum algorithm in~\cite{STV11} for preparing PEPS.}

Before describing in more details how quantum rejection sampling can be used to design a new quantum Metropolis algorithm, let us recall how the standard (classical) Metropolis algorithm works~\cite{metropolis53}. The goal is to solve the following problem: given a classical Hamiltonian associating energies $E_{j}$ to a set of possible configurations $j$, sample from the Gibbs distribution $p(j)=\exp(-\beta E_j)/Z(\beta)$, where $\beta$ is the inverse temperature and $Z(\beta)=\sum_j\exp(-\beta E_j)$ is the partition function. Since the size of the configuration space is exponential in the number of particles, estimating the Gibbs distribution itself is not an option, therefore the Metropolis algorithm proposes to solve this problem by setting up a random walk on the set of configurations that converges to the Gibbs distribution. More precisely, the random walk works as follows:
\begin{enumerate}
\item If $i$ is the current configuration with energy $E_i$, choose a random move to another configuration $j$ (\eg, for a system of spins, a random move could consist in flipping a random spin), and compute the associated energy $E_{j}$.
\item The random move is then accepted or rejected according to the following rule:
\begin{itemize}
  \item if $E_j \leq E_i$, then the move is always accepted;
  \item if $E_j   >  E_i$, then the move is only accepted
        with probability $\exp \bigl( \beta(E_i - E_j) \bigr)$.
\end{itemize}
\end{enumerate}
It can be shown that this random walk converges to the Gibbs distribution.

The quantum Metropolis sampling algorithm by Temme~\textit{et al.}~\cite{TOV+:2009} follows the same general lines as the classical algorithm. It aims at solving the equivalent problem in the quantum case, where we need to generate the thermal state of a Hamiltonian $H$, that is, we need to generate a random eigenstate $\ket{\psi_j}$ where $j$ is sampled according to the Gibbs distribution. The fact that the Hamiltonian is quantum, however, adds a few obstacles, since the set of eigenstates $\ket{\psi_j}$ is not known to start with. The main tool to overcome this difficulty is to use quantum phase estimation~\cite{Kitaev1995,CEMM98} which, applied on the Hamiltonian $H$, allows to project any state on an eigenstate $\ket{\psi_j}$, while obtaining an estimate of the corresponding eigenenergy $E_j$. Similarly to the previous section, we will assume for simplicity that this can be done exactly, that is, we have access to a quantum circuit that acts in the eigenbasis of $H$ as $E_H: \ket{\psi_j} \0{} \mapsto \ket{\psi_j} \ket{E_j}$, where $\ket{E_j}$ exactly encodes the eigenenergy $E_j$. We will also assume that the eigenenergies of $H$ are nondegenerate, so that each eigenenergy $E_j$ corresponds to a single eigenstate $\ket{\psi_j}$, instead of a higher dimensional eigenspace. The quantum Metropolis sampling algorithm also requires to choose a set of quantum gates $\CC$ that will play the role of the possible random moves between eigenstates. In this case, a given quantum gate $C_l\in\CC$ will not simply move an initial eigenstate $\ket{\psi_i}$ to another eigenstate $\ket{\psi_j}$, but rather to a superposition $C_l \ket{\psi_i} = \sum_j c_{ij}^{(l)} \ket{\psi_j}$ where $c_{ij}^{(l)} := \bra{\psi_j} C_l \ket{\psi_i}$.

We can now give a high-level description of the quantum Metropolis sampling algorithm by Temme~\textit{et al.}~\cite{TOV+:2009}. Let $\ket{\psi_i}\ket{E_i}$ be an initial state, that can be prepared by applying the phase estimation operator $E_H$ on an arbitrary state, and measuring the energy register. The algorithm implements each random move by performing the following steps:
\begin{enumerate}
  \item\label{step:qms-random-move}
    Apply a random gate $C_l \in \CC$ on the first register to prepare the state
    $\bigl( C_l \ket{\psi_i} \bigr) \ket{E_i} = \sum_j c_{ij}^{(l)} \ket{\psi_j} \ket{E_i}$.
  \item Apply the phase estimation operator $E_H$ on the $\ket{\psi_j}$ register
    and an ancilla register initialized in the default state $\0{}$ to prepare the state
    $\sum_j c_{ij}^{(l)} \ket{\psi_j} \ket{E_i} \ket{E_j}$.
  \item\label{step:qms-coin}
    Add another ancilla qubit prepared in the state $\ket{0}$ and apply a
    controlled-rotation on this register to create the state
    $\sum_j c_{ij}^{(l)} \ket{\psi_j} \ket{E_i} \ket{E_j}
     \left[ \sqrt{f_{ij}} \ket{1} + \sqrt{1-f_{ij}} \ket{0} \right]$,
    where $f_{ij} := \min \set{1, \exp(\beta(E_i-E_j))}$.
  \item\label{step:qms-reject}
    Measure the last qubit. If the outcome is $0$, reject the move by reverting the
    state to $\ket{\psi_i} \ket{E_i}$ (see~\cite{TOV+:2009} for details) and go
    back to Step~\ref{step:qms-random-move}. Otherwise, continue.
  \item\label{step:qms-end}
    Discard the $\ket{E_i}$ register and measure the $\ket{E_j}$ register
    to project the state onto a new eigenstate $\ket{\psi_j} \ket{E_j}$.
\end{enumerate}
It is shown in~\cite{TOV+:2009} that by choosing a universal set of quantum gates for the set of moves $\CC$, the algorithm simulates random walk on the set of eigenstates of $H$ that satisfies a quantum detailed balanced condition, which ensures that the walk converges to the Gibbs distribution, as in the classical case.

For a given initial state $\ket{\psi_i}\ket{E_i}$, the probability (over all choices of the randomly chosen gate $C_l$) that the measurement in Step~\ref{step:qms-reject} succeeds is $\frac{1}{\abs{\CC}} \sum_{j,l} f_{ij} \abs{c_{ij}^{(l)}}^2$. If we define a vector $\vc{w}^{(i)}$ whose components are $w_{jl}^{(i)} := \sqrt{f_{ij}/\abs{\CC}} c_{ij}^{(l)}$, then this probability is simply $\|\vc{w}^{(i)}\|_2^2$. Hence, after one execution of the algorithm (Steps~\ref{step:qms-random-move}-\ref{step:qms-end}) the initial state $\ket{\psi_i} \ket{E_i}$ gets mapped to $\ket{\psi_j}\ket{E_j}$ with probability $\sum_l \abs{w_{jl}^{(i)}}^2 / \|\vc{w}^{(i)}\|_2^2$. We could achieve the same random move by converting the initial state $\ket{\psi_i}\ket{E_i}$ to
\begin{multline}
   \sum_{j,l} \frac{w_{jl}^{(i)}}{\|\vc{w}^{(i)}\|_2} \ket{l} \ket{\psi_j} \ket{E_i} \ket{E_j} \\
 = \frac{1}{\|\vc{w}^{(i)}\|_2} \sum_j \sqrt{\frac{f_{ij}}{\abs{\CC}}}
   \biggl[ \sum_l c_{ij}^{(l)} \ket{l} \biggr] \ket{\psi_j} \ket{E_i} \ket{E_j}
   \label{eq:target state}
\end{multline}
and discarding the $\ket{l}$ and $\ket{E_i}$ registers and measuring the $\ket{E_j}$ register to project on the state $\ket{\psi_j}\ket{E_j}$ with the correct probability. This implies that one random move reduces to a quantum state conversion problem that we will call $\qmm$. This problem assumes that we are able to perform a perfect phase estimation on the Hamiltonian $H$. Therefore, similarly to the previous section, we fix a set of possible eigenenergies $\EE$ of finite cardinality $n := \abs{\EE}$.

\begin{definition}\deftitle{Quantum Metropolis move}
The \emph{quantum Metropolis move problem}, denoted by $\qmm$, is a quantum state conversion problem $(\O, \Phi, \Psi, \X)$, where
\begin{equation*}
  \X := \set{(H,i) \in \Herm(d) \times [d]: \spec(H) \in \EE^d}.
\end{equation*}
Oracles in $\O$ act as $E_H: \ket{\psi_j} \0{n} \mapsto \ket{\psi_j} \ket{E_j}$, with the corresponding initial states $\ket{\psi_i}$, the eigenvectors of $H$, and target states $\sum_{j,l} w_{jl}^{(i)} / \|\vc{w}^{(i)}\|_2 \ket{l} \ket{\psi_j}$ where
$w_{jl}^{(i)} := \sqrt{f_{ij}/\abs{\CC}} c_{ij}^{(l)}$,
$f_{ij}       := \min \set{1, \exp(\beta(E_i-E_j))}$, and
$c_{ij}^{(l)} := \bra{\psi_j} C_l \ket{\psi_i}$.
\label{def:qmm}
\end{definition}

A critical part of the algorithm from~\cite{TOV+:2009} described above is how to revert a rejected move in Step~\ref{step:qms-reject}. Temme~\textit{et al.} show how this can be done by using an unwinding technique based on successive measurements, but we will not describe this technique in detail, as we now show how this step can be avoided by using quantum rejection sampling. Intuitively, this can be done by using amplitude amplification to ensure that the measurement in Step~\ref{step:qms-reject} always projects on the ``accept'' state $\ket{1}$. This also avoids having to repeatedly attempt random moves until one is accepted, and the number of steps of amplitude amplification will be quadratically smaller than the number of random moves that have to be attempted until one is accepted. This leads to the following statement:
\thmmetropolis*
\begin{proof}
The modified algorithm follows the same lines as the original algorithm, except that Steps~\ref{step:qms-coin}-\ref{step:qms-reject} are replaced by a quantum rejection sampling step. We use a quantum coin to choose the random gate in Step~\ref{step:qms-random-move} in order to make it coherent. The algorithm starts by applying the phase estimation oracle $E_H$ on the initial state to prepare the state $\ket{\psi_i} \ket{E_i}$, and then proceeds with the following steps:
\begin{enumerate}
  \item\label{step:qms2-random-move}
    Prepare an extra register in the state $\frac{1}{\sqrt{\abs{\CC}}}\sum_l\ket{l}$.
    Conditionally on this register, apply the gate $C_l \in \CC$
    on the eigenstate register to prepare the state
    \begin{equation*}
      \frac{1}{\sqrt{\abs{\CC}}} \sum_j
      \left[ \sum_l c_{ij}^{(l)} \ket{l} \right]
      \ket{\psi_j} \ket{E_i}.
    \end{equation*}
  \item\label{step:qms2-pe}
    Apply the phase estimation operator $E_H$ on the second register and an ancilla
    register initialized in the default state $\0{n}$ to prepare the state
    \begin{equation*}
      \frac{1}{\sqrt{\abs{\CC}}} \sum_j
      \left[ \sum_l c_{ij}^{(l)} \ket{l} \right]
      \ket{\psi_j} \ket{E_i} \ket{E_j}.
    \end{equation*}
  \item\label{step:qms2-sqrs}
    Convert this state to the state given in Eq.~(\ref{eq:target state}):
    \begin{equation*}
      \frac{1}{\|\vc{w}^{(i)}\|_2} \sum_j
      \sqrt{\frac{f_{ij}}{\abs{\CC}}}
      \left[ \sum_l c_{ij}^{(l)} \ket{l} \right]
      \ket{\psi_j} \ket{E_i} \ket{E_j}.
    \end{equation*}
  \item
    Discard $\ket{E_i}$ and uncompute $\ket{E_j}$
    by using one call to the phase estimation oracle $E_H^\dagger$.
\end{enumerate}

Note that Step~\ref{step:qms2-sqrs} is an instance of $\sqsamplingab$, where the pair of basis states $\ket{E}\ket{E'}$ of the phase estimation registers plays the role of the states $\ket{k}$, the initial amplitudes $\pi_{E,E'}$ are given by $\frac{1}{\sqrt{\abs{\CC}}}\sqrt{\sum_l\abs{c_{ij}^{(l)}}^2}$ for $(E,E')=(E_i,E_j)$ or $0$ for values $(E,E')$ that do not correspond to a pair of eigenvalues of $H$, the states $\left[ \sum_l c_{ij}^{(l)} \ket{l} \right] \ket{\psi_j} / \sqrt{\sum_l\abs{c_{ij}^{(l)}}^2}$ play the role of the unknown states $\ket{\xi_k}$, and the ratio between the initial and target amplitudes is given by $\tau_{E,E'} = \sqrt{\min \set{1, \exp(\beta(E-E'))}}$ (the reduction is summarized in Table~\ref{table:reduction-qms}). Therefore, this step may be performed using the algorithm in Lemma~\ref{lem:algo-sqsqmpling}. Here, we choose $\alpha=1$ since the full Quantum Metropolis Sampling algorithm requires to apply a large number of successive random moves, therefore each instance of $\qmm$ should be solved with high success probability. Choosing $\alpha=1$ ensures that each random move will have success probability 1 (under our assumption that the phase estimation oracle is perfect), using an expected number of phase estimation oracles $O(1/\|\vc{w}^{(i)}\|_2)$.
\qed
\end{proof}


Note that in this case it is critical that the algorithm only requires one copy of the initial state, hence solving the quantum state conversion problem $\sqsamplingab$ (in contrast, the quantum algorithm for linear systems of equations used a unitary to create multiple copies of the initial state, which is allowed only in the weaker quantum state generation problem $\qsamplingab$). Indeed, creating the initial state requires one copy of the previous eigenstate $\ket{\psi_i}$, which cannot be cloned as it is unknown. Here, the algorithm only requires to reflect through the initial state, which can be done by inverting Steps~\ref{step:qms2-random-move}-\ref{step:qms2-pe}, applying a phase $-1$ conditionally on the eigenenergy register being in the state $\ket{E_i}$ (which is possible since $E_i$ is known), and applying Steps~\ref{step:qms2-random-move}-\ref{step:qms2-pe} again.

Repeating the algorithm for $\qmm$ a large number of times will simulate the same random walk on the eigenstates of $H$ as the original quantum Metropolis sampling algorithm in~\cite{TOV+:2009}, except that we have a quadratic speed-up over the number of attempted moves necessary to obtain an accepted move. In order to converge to the Gibbs distribution, we need to take into account this quadratic speed-up in order to decide when to stop the algorithm, effectively assuming that each move takes quadratically longer than it actually does. Another option would be to modify the algorithm for $\qsamplingab$ so that it also estimates $\|\vc{w}^{(i)}\|_2$ by using amplitude estimation or quantum counting~\cite{BoyerBHT98,AmplitudeAmplification}. We leave the full analysis of these technical issues for future work.

\begin{table*}
\begin{center}
\begin{tabular}{|c|c|}
\hline
$\sqsamplingab$ & $\qmm$ \\
\hline
$\ket{k}$ & $\ket{E}\ket{E'}$\\
$\pi_k$ & $\pi_{E,E'}:=
\begin{cases}
  \frac{1}{\sqrt{\abs{\CC}}}\sqrt{\sum_l\abs{c_{ij}^{(l)}}^2}
  \textrm{ if } (E,E') = (E_i,E_j) \textrm{ where } E_i, E_j \in \spec(H) \\
  0 \textrm{ if } E \notin \spec(H) \textrm{ or } E' \notin \spec(H)
\end{cases}
$\\
$\ket{\xi_k}$ & $\ket{\xi_{E,E'}}:=
\begin{cases}
  \sum_l c_{ij}^{(l)}\ket{l}\ket{\psi_j}/\sqrt{\sum_l\abs{c_{ij}^{(l)}}^2}
  \textrm{ if } (E,E') = (E_i,E_j) \textrm{ where } E_i, E_j \in \spec(H) \\
  \textrm{n/a if } E \notin \spec(H) \textrm{ or } E' \notin \spec(H)
\end{cases}
$\\
$\tau_k$ & $\tau_{E,E'} := \sqrt{\min \set{1, \exp(\beta(E-E'))}}$ \\
\hline
\end{tabular}
\end{center}
\caption{Reduction from Step~\ref{step:qms2-sqrs} in the new quantum Metropolis sampling algorithm to the quantum resampling problem \textnormal{$\sqsamplingab$}.}
\label{table:reduction-qms}
\end{table*}

\subsection{Boolean hidden shift problem} \label{sect:hiddenshift}



Our final application of the quantum algorithm for the $\qsamplingab$ problem is a new quantum algorithm for the Boolean hidden shift problem $\bhsp_f$.

We refer to the recent review~\cite{DeWolf:2008} for a good overview
of basic properties of the Fourier transform of Boolean functions.
Fourier analysis on the Boolean cube studies the $2^n$-dimensional
vector space of all \emph{real}-valued functions defined on the
$n$-dimensional Boolean cube $\bb{n}$. Thus, in the following definition $f$
denotes a function of the form $\bb{n} \to \R$ (\emph{not} a
Boolean function). The Boolean case is discussed later.


\begin{definition}\deftitle{Fourier transform}
The \emph{Fourier basis} of $\bb{n}$ consists of functions $\set{\chi_{\bv{w}} : \bv{w} \in \bb{n}}$, where each $\chi_{\bv{w}} : \bb{n} \to \set{1,-1}$ is defined as
$
  \chi_{\bv{w}}(\bv{x}) := (-1)^{\bv{w} \cdot \bv{x}},
$
where $\bv{w} \cdot \bv{x} := \sum_{i=1}^n w_i x_i$ is the inner product in $\b$. The \emph{Fourier transform} of a function $f : \bb{n} \to \R$ is the function $\hat{f}: \bb{n} \to \R$ defined as
$
  \hat{f}(\bv{w}) := 
\frac{1}{2^n} \sum_{\bv{x} \in \bb{n}} (-1)^{\bv{w} \cdot \bv{x}} f(\bv{x}).
$
Note that $\hat{f}(\bv{w}) := \E_{\bv{x}}(\chi_{\bv{w}} f) = \frac{1}{2^n}
\sum_{\bv{x} \in \bb{n}} \chi_{\bv{w}}(\bv{x}) f(\bv{x})$ which is another way to write the
Fourier coefficients.
The set $\set{\hat{f}(\bv{w}) : \bv{w} \in
  \bb{n}}$ of all values of $\hat{f}$ is called the \emph{spectrum} of
$f$ and each of its elements is called a \emph{Fourier coefficient} of
$f$.
\end{definition}



Let us consider a \emph{Boolean function} \mbox{$f: \bb{n} \to \b$}. To find its Fourier transform, it is
required to associate $f$ with a real-valued function \mbox{$F : \bb{n} \to \R$} in some way. Instead of the obvious correspondence (treating $0,1 \in \b$ as real numbers) for the purposes of this work it is more natural to let $F$ be the $(\pm 1)$-valued function defined by $F(\bv{x}) := (-1)^{f(\bv{x})}$.

\begin{definition}\deftitle{Fourier transform of a Boolean function}
By slight abuse of notation, the \emph{Fourier transform of a
Boolean function} \mbox{$f: \bb{n} \to \b$} is the function
$\hat{f}: \bb{n} \to \R$ defined as
\begin{equation*}
  \hat{f}(\bv{w})
  := \E_{\bv{x}}(\chi_{\bv{w}} F)
  = \frac{1}{2^n} \sum_{\bv{x} \in \bb{n}} (-1)^{\bv{w} \cdot \bv{x} + f(\bv{x})}.
\end{equation*}
\end{definition}

Based on a reduction to $\qsamplingab$, we can now prove the following upper bound on the query complexity of $\bhsp_f$.

\thmBHSP*

\begin{proof}
We will use $\oracle{\bv{s}}$ to denote the \emph{phase oracle} for function $\fs$, \ie, a diagonal matrix that acts on the standard basis vectors $\bv{x} \in \bb{n}$ as follows:
$\oracle{\bv{s}} \ket{\bv{x}} := (-1)^{f(\bv{x}+\bv{s})} \ket{\bv{x}}$.
Let us consider the quantum oracle $\oracle{\bv{s}}$ conjugated by the Hadamard transform. The resulting operation
\begin{equation*}
  \Vs := \H{n} \, \oracle{\bv{s}} \, \H{n}
\end{equation*}
is very useful, since, when acting on a register initialized in all-zeros state, it can be used to prepare the following quantum superposition:
\begin{equation*}
  \ket{\psifs}
  := \Vs \ket{0}\xp{n}
   = \sum_{\bv{w} \in \bb{n}} (-1)^{\bv{w} \cdot \bv{s}} \hat{f}(\bv{w}) \ket{\bv{w}}.
\end{equation*}

If we could eliminate the Fourier coefficients $\hat{f}(\bv{w})$ from state $\ket{\psifs}$, we would obtain a state
\begin{equation*}
  \ket{\psis} := \frac{1}{\sqrt{2^n}} \sum_{\bv{w} \in \bb{n}} (-1)^{\bv{w} \cdot \bv{s}} \ket{\bv{w}}
\end{equation*}
from which the hidden shift $\bv{s}$ can be easily recovered by applying the Hadamard transform $\H{n}$.
Luckily, the problem of transforming the state $\ket{\psifs}$ to $\ket{\psis}$ is a special case of $\qsamplingab$ with
\begin{align*}
  \quantump_{\bv{w}} &:= \abs{\hat{f}(\bv{w})}, &
  \quantums_{\bv{w}}  &:= 1/\sqrt{2^n}, &
  \ket{\xi_{\bv{w}}} &:= (-1)^{\bv{w} \cdot \bv{s}} \ket{0}.
\end{align*}
(More precisely, the initial amplitudes are $\hat{f}(\bv{w})$ instead of $\abs{\hat{f}(\bv{w})}$. However, the function $f$ and therefore its Fourier transform $\hat{f}$ is completely known, so we can easily correct the phases using a controlled-phase gate.) As a consequence, Theorem~\ref{thm:qsampling} immediately gives us a quantum algorithm for solving this problem.
\qed
\end{proof}

The complexity of the algorithm is limited by the smallest Fourier coefficient of the function. By ignoring small Fourier coefficients, one can decrease the complexity of the algorithm, at the cost of a lower success probability. However, the success probability of this algorithm can be boosted using repetitions, which requires to construct a procedure to check a candidate shift. We propose such a checking procedure based on a controlled-SWAP test. The number of necessary repetitions may then be decreased quadratically using the amplitude amplification technique of~\cite{HMW03} (note that we cannot use the usual amplitude amplification algorithm since the checking procedure is imperfect). This leads to the following theorem (proved in Appendix~\ref{sect:boosting}):
\thmBHSPboost*

\section*{Conclusion and open problems}

We provide an algorithm for solving the quantum resampling problem. Our algorithm can be viewed as a quantum version of the classical rejection sampling technique.
It relies on amplitude amplification~\cite{AmplitudeAmplification} to increase the amplitude of some target ``accept'' state, and its query complexity is given by a semidefinite program.
The solution of this SDP and hence the cost of the algorithm depends on the ratio between the amplitudes of the initial and target states, similarly to the case of the classical rejection sampling where the cost is given by the ratio of probabilities. Using the automorphism principle over a unitary group, we derive an SDP for the lower bound that is identical to the one for the upper bound, showing that our algorithm has optimal query complexity.

While the original adversary method cannot be applied as is for this quantum state generation problem because the oracle encodes an unknown quantum state instead of some unknown classical data, it is interesting to note that the query complexity of this problem is also characterized by an SDP. Therefore, an interesting open question is whether the adversary method~\cite{Amb00,NegativeWeights}, which has been shown to be tight for evaluating functions~\cite{Reichardt:2009,ReichardtReflections,LMRSS11} and nearly tight for quantum state generation or conversion problems with classical oracles~\cite{LMRSS11}, can always be extended and shown to be tight for this more general framework of problems with quantum oracles.

In Sect.~\ref{sect:applications}, we illustrate how quantum rejection sampling may be used as a primitive in algorithm design by providing three different applications. We first show that it was used implicitly in the quantum algorithm for linear systems of equations~\cite{HHL:09}. By assuming a perfect phase estimation operator on the matrix of the system, we show that this problem reduces to a quantum state conversion problem which we call $\qle$, which itself reduces to $\sqsamplingab$. An open question is how to combine the quantum rejection sampling approach with the variable time amplitude amplification technique that was proposed by Ambainis~\cite{Ambainis2010} to improve on the original algorithm by Harrow \textit{et al.}~\cite{HHL:09}. In order to do so, we should ``open'' the phase estimation black box since Ambainis's main idea is to stop some branches of the phase estimation earlier than others.

As a second application, we show that quantum rejection sampling can be used to speed up the main step in the original quantum Metropolis sampling algorithm~\cite{TOV+:2009}. The general idea is to use amplitude amplification to increase the acceptance probability of a move, and therefore quadratically reduce the number of moves that have to be attempted before one is accepted. While this approach also provides some type of quadratic speed-up, it is rather different from the ``quantum-quantum'' Metropolis algorithm proposed by Yung and Aspuru-Guzik~\cite{YA:2010}. The main difference is that the approach based on quantum rejection sampling still simulates the same classical random walk on the eigenstates of the Hamiltonian, whereas the quantum-quantum Metropolis algorithm replaces it by a quantum walk. Note that while random walks converge towards their stationary distribution from any initial state, this is not the case for quantum walks as they are reversible by definition. Therefore, while both the original quantum Metropolis sampling algorithm and our variation can start from any initial state and run at a fixed inverse temperature $\beta$ to converge to the corresponding Gibbs distribution, the quantum-quantum Metropolis sampling algorithm works differently: it starts from a uniform superposition, which corresponds to the Gibbs distribution at $\beta=0$, and uses a series of measurements to project this state onto superpositions corresponding to Gibbs distributions with increasingly large $\beta$, until the desired value is reached.

Finally. as shown in Sect.~\ref{sect:hiddenshift}, we can apply the quantum rejection sampling technique to solve the hidden shift problem for any Boolean function $f$. In the limiting cases of flat or highly peaked Fourier spectra we recover the quantum algorithm for bent functions~\cite{Roetteler:2010} or Grover's algorithm for delta functions~\cite{Grover:96}, respectively. For a general Boolean function the hidden shift problem can be seen as lying somewhere between these two extreme cases. While the algorithm is known to be optimal for the extreme cases of bent and delta functions, its optimality for more general cases remains an open problem. A related question is the optimality of the checking procedure that leads to Theorem~\ref{thm:BHSP2}.

\section*{Acknowledgements}

The authors acknowledge support by ARO/NSA under grant W911NF-09-1-0569
and also wish to thank Andrew Childs, Dmitry Gavinsky, Sean Hallgren and Guosen Yue for fruitful discussions. M.O. acknowledges support from QuantumWorks and the US ARO/DTO and would like to thank Martin Schwarz and Kristan Temme for fruitful discussions.


\bibliographystyle{abbrvurl}
\bibliography{qrs-itcs}

\appendix

\section{Water-filling vector is\\optimal for the SDP} \label{apx:SDP}

\lemSDPsolution*

\begin{proof}
We now show that the optimal value of the SDP in Eq.~(\ref{eq:Primal}) can be attained by a rank-$1$ matrix $M$. Imposing the additional constraint that $M$ can be written as $M = \ve \cdot \ve\tp$ for some $\ve \in \R^n$, the optimization problem~(\ref{eq:Primal}) reduces to
\begin{equation}
\begin{array}{rl}
  \max_{\e_k \geq 0} \norm{\ve}_2^2
  \quad \text{s.t.} \quad
  & \forall k: \quantump_k \geq \e_k \geq 0, \\
  & \vc{\quantums}\tp \cdot \hat{\ve} \geq \sqrt{\pp},
\end{array}
\label{eq:Primal1}
\end{equation}
where $\hat{\ve} := \ve / \norm{\ve}_2$ denotes a unit vector in direction $\ve$.

We show that the optimal value is attained by $\ve=\waterfilling$. Recall that by Def.~\ref{def:water-filling}, we have $\e_k=\min \set{\quantump_k, \gamma \quantums_k}$ and
\begin{equation}\label{eq:saturate-p}
  \vc{\quantums}\tp \cdot \hat{\ve} = \sqrt{\pp},
\end{equation}
so that this vector satisfies the constraints in~(\ref{eq:Primal1}) and is therefore a feasible point. As a consequence $M=\ve \cdot \ve\tp$ is also a feasible point for the SDP~(\ref{eq:Primal}), which implies that its objective value is at least $\Tr M=\norm{\ve}_2^2$.

We now want to find a feasible dual solution that gives the same objective value for the dual of SDP~(\ref{eq:Primal}), which can be written as~\cite{VB:96}
\begin{equation}
  \min_{\lambda_k \geq 0, \, \mu \geq 0} \sum_{k=1}^n \lambda_k \quantump_k^2
  \quad \text{s.t.} \quad
   \Lambda - I + \mu (\pp I - \vc{\quantums} \cdot \vc{\quantums}\tp) \sgeq 0,
\label{eq:Dual}
\end{equation}
where $\Lambda := \diag(\lambda_k \mid k = 1, \dotsc, n)$. Indeed, if an objective value is feasible for both the primal and the dual, it implies that this is the optimal value.


We prove that the following solution is feasible for the dual:
\begin{align}
  \lambda_k &= \mu \biggl( \frac{\quantums_k}{\e_k}
               \sum_{l=1}^n \quantums_l \e_l - \pp \biggr) + 1, \nonumber \\
  \mu &= \frac{1 - \norm{\ve}_2^2}{\pp-\bigl(\sum_{l=1}^n \quantums_l \e_l\bigr) \cdot
         \bigl( \sum_{k=1}^n\frac{\quantums_k\quantump_k^2}{\e_k}\bigr)}.
  \label{eq:lambda mu}
\end{align}
This choice yields $\norm{\ve}_2^2$ as the dual objective value, so it remains to show that it satisfies the constraints in~(\ref{eq:Dual}).
Let us first prove that $\mu \geq 0$, which is equivalent to
\begin{equation}\label{eq:condition-mu}
\biggl(\sum_{l=1}^n \quantums_l \e_l\biggr)\cdot\biggl( \sum_{k=1}^n\frac{\quantums_k\quantump_k^2}{\e_k}\biggr) \leq \pp.
\end{equation}
Let us decompose the vector $\vc{\quantump}$ into two orthogonal parts such that $\vc{\quantump}=\vc{\quantump}_\leq+\vc{\quantump}_>$, where $\vc{\quantump}_\leq$ corresponds to components $\quantump_k$ such that $\quantump_k\leq\gamma\quantums_k$, and $\vc{\quantump}_>$ to the remaining components. Decomposing $\vc{\quantums}$ and $\ve$ similarly, we have $\ve=\vc{\quantump}_\leq+\gamma\vc{\quantums}_>$. The following are straightforward
\begin{align*}
 1&=\norm{\vc{\quantump_\leq}}_2^2+\norm{\vc{\quantump_>}}_2^2\\
\norm{\ve}_2^2&=\norm{\vc{\quantump_\leq}}_2^2+\gamma^2\norm{\vc{\quantums_>}}_2^2\\
\ve\tp\cdot\vc{\quantums}&=\vc{\quantump\tp_\leq}\cdot\vc{\quantums_\leq}+\gamma\norm{\vc{\quantums_>}}_2^2.
\end{align*}
Using these equalities, we obtain
\begin{align}
 \sum_{k=1}^n\frac{\quantums_k\quantump_k^2}{\e_k}&=\vc{\quantump\tp_\leq}\cdot\vc{\quantums_\leq}+\frac{1}{\gamma}\norm{\vc{\quantump_>}}_2^2\nonumber\\
&=\vc{\quantump\tp_\leq}\cdot\vc{\quantums_\leq}+\frac{1}{\gamma}\left( 1-\norm{\vc{\quantump_\leq}}_2^2\right)\nonumber\\
&=\ve\tp\cdot\vc{\quantums}+\frac{1}{\gamma}\left( 1-\norm{\ve}_2^2\right).
\label{eq:rhs-condition-mu}
\end{align}
Therefore, the left hand side of~(\ref{eq:condition-mu}) can be written as
\begin{align*}
\biggl(\sum_{l=1}^n \quantums_l \e_l\biggr)\cdot\biggl( \sum_{k=1}^n\frac{\quantums_k\quantump_k^2}{\e_k}\biggr)
&=\left(\ve\tp\cdot\vc{\quantums}\right)^2
 +\frac{\ve\tp\cdot\vc{\quantums}}{\gamma}\left( 1-\norm{\ve}_2^2\right)\\
&=\pp\norm{\ve}_2^2+\frac{\ve\tp\cdot\vc{\quantums}}{\gamma}\left( 1-\norm{\ve}_2^2\right)\\
&=\left(\pp-\frac{\ve\tp\cdot\vc{\quantums}}{\gamma}\right)\norm{\ve}_2^2+\frac{\ve\tp\cdot\vc{\quantums}}{\gamma},
\end{align*}
where we have used~(\ref{eq:saturate-p}). Since $\e_k\leq\gamma\quantums_k$, we have $\norm{\ve}_2^2\leq \gamma\ve\tp\cdot\vc{\quantums}$, which, together with~(\ref{eq:saturate-p}) implies that 
$
\frac{\ve\tp\cdot\vc{\quantums}}{\gamma}\leq \pp.
$
Together with $\norm{\ve}_2^2\leq\norm{\vc{\quantump}}_2^2=1$, this implies
\begin{align*}
\left(\pp-\frac{\ve\tp\cdot\vc{\quantums}}{\gamma}\right)\norm{\ve}_2^2+\frac{\ve\tp\cdot\vc{\quantums}}{\gamma}
\leq \pp,
\end{align*}
which proves~(\ref{eq:condition-mu}) and, in turn, $\mu\geq 0$.

We now show that $\lambda_{k} \geq 0$ for all $k\in[n]$. From Eqs.~(\ref{eq:lambda mu}) we see that this is equivalent to showing
\begin{equation*}
\frac{\quantums_k}{\e_k} \ve\tp\cdot\vc{\quantums} - \pp
\geq -\frac{\pp-\ve\tp\cdot\vc{\quantums}\sum_{k=1}^n\frac{\quantums_k\quantump_k^2}{\e_k}}{1 - \norm{\ve}_2^2}
\end{equation*}
Note that $1 \geq \norm{\ve}_2^2$. By multiplying out everything with $1 - \norm{\ve}_2^2$ and expanding, we get
\begin{equation*}
  \frac{\quantums_k}{\e_k} \ve\tp\cdot\vc{\quantums}\bigl( 1 - \norm{\ve}_2^2 \bigr)
  + \pp \norm{\ve}_2^2
  \geq \ve\tp\cdot\vc{\quantums}\sum_{k=1}^n\frac{\quantums_k\quantump_k^2}{\e_k}.
\end{equation*}
Note that $\pp \norm{\ve}_2^2 = (\ve\tp\cdot\vc{\quantums})^2$, so after rearranging terms and dividing by $\ve\tp\cdot\vc{\quantums}$ we get
\begin{equation*}
\frac{\quantums_k}{\e_k} \bigl( 1 - \norm{\ve}_2^2 \bigr)
+\ve\tp\cdot\vc{\quantums}\geq \sum_{k=1}^n\frac{\quantums_k\quantump_k^2}{\e_k}.
\end{equation*}
We apply Eq.~(\ref{eq:rhs-condition-mu}) to the right hand side and get
\begin{equation*}
\frac{\quantums_k}{\e_k} \bigl( 1 - \norm{\ve}_2^2 \bigr)
+\ve\tp\cdot\vc{\quantums}\geq\ve\tp\cdot\vc{\quantums}
 + \frac{1}{\gamma} \bigl( 1 - \norm{\ve}_2^2 \bigr).
\end{equation*}
After simplification this yields $\e_k\leq\gamma\quantums_k$, which is true by definition of $\ve$. Thus, we have $\lambda_{k} \geq 0$.

Finally, it remains to show that the following matrix is positive semidefinite:
\begin{equation*}
  \Lambda - I + \mu (\pp I - \vc{\quantums} \cdot \vc{\quantums}\tp)\! =\! \mu\! \left[\biggl(\sum_{l=1}^n \quantums_l \e_l\biggr)\cdot\diag(\quantums_k/\e_k)-\vc{\quantums} \cdot \vc{\quantums}\tp\right].
\end{equation*}
Since $\mu\geq 0$, it is the case if and only if
\begin{equation*}
  \forall \vc{v} \in \R^{n}: \quad
\biggl(\sum_{l=1}^n \quantums_l \e_l\biggr)\cdot\biggl(\sum_{k=1}^n \frac{v_k^2\quantums_k}{\e_k}\biggr) \geq \biggl(\sum_{k=1}^n v_k\quantums_k\biggr)^2.
\end{equation*}
This follows by Cauchy--Schwarz inequality:
$\bigl( \sum_{l=1}^n \quantums_l \e_l \bigr) \cdot
 \bigl( \sum_{k=1}^n v_k^2\quantums_k/\e_k \bigr) \geq
 \bigl( \sum_{k=1}^n \sqrt{\quantums_k\e_k} \cdot \sqrt{v_k^2\quantums_k/\e_k} \bigr)^2$.
\qed
\end{proof}

\section{Boosting the success probability}\label{sect:boosting}

When we want to find the hidden shift with probability close to one, the algorithm in Theorem~\ref{thm:BHSP} might be quite inefficient as, in the special case $p=1$, its complexity is limited by the smallest Fourier coefficient (in particular, when some Fourier coefficients are zero, it is not possible to obtain $p=1$). Another approach to find the hidden shift with near certainty would be to use the algorithm with a smaller $p$ and repeat until the right shift is obtained. However, this requires a procedure to check a candidate shift. Classically this can be done, \eg, by querying the oracle on several inputs uniformly at random and checking if the output agrees with the shifted function. Using this strategy one can boost the success probability arbitrarily close to $1$ by using $O(1/\gf)$ queries to the shifted function $\fs$.

\begin{definition}\deftitle{Influence}
For any Boolean function $f$ and $\bv{v}\in\bb{n}$, we call  $\gv = \Pr_x[f(\bv{x}) \neq f(\bv{x}+\bv{v})]$ the \emph{influence} of $\bv{v}$ over $f$, and $\gf=\min_{\bv{v}}\gv$ the \emph{minimum influence} over $f$.
\end{definition}

\noindent However, on a quantum computer one can check a candidate shift with only $O(1/\sqrt{\gf})$ oracle calls using a procedure similar to controlled-SWAP test. If we combine this imperfect checking procedure with the quantum search algorithm with bounded-error inputs from~\cite{HMW03}, we get the following result:

\thmBHSPboost*

\begin{figure}[th]
  
  \centering
  \TikZorPDF{
    \input{fig-checking.tex}
  }{
    \includegraphics[width=0.35\textwidth]{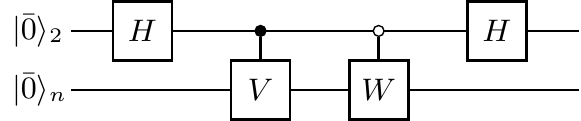}
  }

  \caption{A one-sided test for checking if states $\ket{\psi}$ and $\ket{\phi}$ are equal. If $V \0{n} = \ket{\psi}$ and $W \0{n} = \ket{\varphi}$ then the above circuit prepares a state $\frac{1}{2} \bigl[ \ket{0} (\ket{\psi} + \ket{\varphi}) + \ket{1} (\ket{\psi} - \ket{\varphi}) \bigr]$. If the first register is measured in the standard basis, we get outcome $0$ (``accept'') with probability $p(0) = \frac{1}{4} \norm{\ket{\psi} + \ket{\varphi}}_2^2 = \frac{1}{2} (1 + \Re \braket{\psi}{\varphi})$. Note that we never get outcome $1$ if $\ket{\psi} = \ket{\varphi}$.}
  \label{fig:Checking}
\end{figure}

\begin{proof}
Let us first show how to check a candidate shift $\bv{v}$, assuming that $\bv{s}$ is the actual shift. Using one call to the oracle $\oracle{\bv{s}}$, we can easily prepare the state
\begin{align*}
 \ket{\phi_f(\bv{s})}&=\frac{1}{\sqrt{2^n}}\sum_{\bv{x}}(-1)^{f(\bv{x}+\bv{s})}\ket{\bv{x}}
\end{align*}
Similarly, we can prepare the state $\ket{\phi_f(\bv{v})}$ by applying a transformation corresponding to oracle $\oracle{\bv{v}}$. The inner product between these two states is
\begin{align*}
 \braket{\phi_f(\bv{s})}{\phi_f(\bv{v})}&=\frac{1}{2^n}\sum_{\bv{x}}(-1)^{f(\bv{x}+\bv{s})+f(\bv{x}+\bv{v})}\\
&=1-2\gf(\bv{s}+\bv{v}).
\end{align*}
Therefore, the procedure from Fig.~\ref{fig:Checking} will accept the candidate shift $\bv{v}$ with probability
\begin{align*}
 \frac{1+\braket{\phi_f(\bv{s})}{\phi_f(\bv{v})}}{2}=1-\gf(\bv{s}+\bv{v}),
\end{align*}
that is, it will always accept if $\bv{s}=\bv{v}$, and reject with probability at least $\gf$ otherwise. Repeating this test $O(1/\gf)$ times, we can ensure that a wrong candidate shift is rejected with high probability. Moreover, using once again quantum amplitude amplification~\cite{AmplitudeAmplification}, we can obtain a quadratic improvement, and therefore reject a wrong candidate using only $O({1}/{\sqrt{\gf}})$ oracle calls.

Using the algorithm from Theorem~\ref{thm:BHSP} for success probability $p$, we can then boost the success probability to any constant $1-\delta$ using the quantum search algorithm with bounded-error inputs of~\cite{HMW03} (note that using the usual quantum amplitude amplification technique~\cite{AmplitudeAmplification} would incur an additional factor of $\log(1/p)$ since the checking operation is imperfect). Therefore, the total complexity comes from repeating $O(1/\sqrt{p})$ times the algorithm in Theorem~\ref{thm:BHSP}, with cost $O(1/\norm{\ve}_2)$, and the checking operation, with cost\linebreak $O(1/\sqrt{\gf})$.
\qed
\end{proof}

\end{document}

%% file: fig-model.tex
\def\h{0.5}
\def\w{0.85}


\newcommand{\ClassicalRejectionSampling}{
\begin{tikzpicture}
  [bbox/.style = {draw = black, fill = black!15, thick, minimum height = 1.6cm, minimum width = 0.65cm, rectangle},
   circ/.style = {draw = black, fill = white,    thick, inner sep = 0.95mm, circle},
   orac/.style = {draw = black, fill = black!15, thick, minimum height = 1.6cm, minimum width = 0.65cm, rectangle},
   gate/.style = {draw = black, fill = white,    thick, minimum height = 0.6cm, minimum width = 0.65cm, rectangle}
  ]
  \def\xmax{3.5*\w}
  \draw [->, thick] (0,+\h) to (\xmax,+\h);
  \draw [->, thick] (0,-\h) to (\xmax,-\h);
  \node (P) at (0, 0)   [bbox] {$P$};
  \node     at (1*\w,+\h) [fill = white] {$\xi(k)$};
  \node     at (1*\w,-\h) [fill = white] {$k$};
  \node (A) at (2*\w,-\h) [circ] {$\mc{A}$};
  \draw [->, thick] (A) -- +(0,\h) -- (\xmax,0);
  \node     at (\xmax+1.5*\w,+\h) {$\xi(k)$};
  \node     at (\xmax+1.5*\w,0) {accept/reject};
  \node     at (\xmax+1.5*\w,-\h) {$k$};
\end{tikzpicture}
}

\ClassicalRejectionSampling


\newcommand{\QuantumRejectionSampling}{
\begin{tikzpicture}
  [bbox/.style = {draw = black, fill = black!15, thick, minimum height = 1.6cm, minimum width = 0.65cm, rectangle},
   circ/.style = {draw = black, fill = white,    thick, inner sep = 0.95mm, circle},
   orac/.style = {draw = black, fill = black!15, thick, minimum height = 1.6cm, minimum width = 0.65cm, rectangle},
   gate/.style = {draw = black, fill = white,    thick, minimum height = 0.6cm, minimum width = 0.65cm, rectangle}
  ]
  \def\x{5};
  \def\W{5.7};
  \draw [-, thick] (\x+0.2,+\h) to (\x+\W,+\h);
  \draw [-, thick] (\x+0.2,-\h) to (\x+\W,-\h);
  \draw (\x+1*\w,  0) node [orac] {$O$};
  \draw (\x+2*\w,-\h) node [gate] {$U_1$};
  \draw (\x+3*\w,  0) node [orac] {$O\ct$};
  \draw (\x+4*\w,-\h) node [gate] {$U_2$};
  \draw (\x+5*\w,  0) node [orac] {$O$};
  \node [fill = white] at (\x+6*\w,-\h) {\ldots};
  \node [fill = white] at (\x+6*\w,+\h) {\ldots};
  \node at (\x+\W+0.3,+\h) {$\ket{\xi_k}$};
  \node at (\x+\W+0.3,-\h) {$\ket{k}$};  
\end{tikzpicture}
}

%% file: fig-problems.tex
\begin{tikzpicture}
  \def\d{0.15}
  \def\t{0.3}
  \def\w{0.6}

  \def\bot{-1}
  \def\top{+2}

  \draw [rounded corners]           (0*\w, \bot+0*\d   ) rectangle (8.5, \top-0*\w);
  \draw node [right] at             (0*\w, \top-0*\w-\t) {Quantum state conversion};

  \draw [rounded corners]           (1*\d, \bot+1*\d   ) rectangle (7.5, \top-1*\w);
  \draw node [right] at             (1*\d, \top-1*\w-\t) {Quantum state generation};

  \draw [rounded corners, black!60] (2*\d, \bot+2*\d   ) rectangle (8.5-\d, \top-2*\w);
  \draw node [right, black!60] at   (2*\d, \top-2*\w-\t) {Classical oracles};

  \draw [rounded corners]           (3*\d, \bot+3*\d   ) rectangle (4, \top-3*\w);
  \draw node [right] at             (3*\d,  2-3*\w-\t  ) {Function evaluation};

  \draw node [left] at ( 8.5, \top-0*\w-\t) {$\bullet\ \sqsamplingab$};
  \draw node [left] at ( 7.5, \top-1*\w-\t) {$\bullet\ \qsamplingab$};
  \draw node [left] at ( 7.5, \top-3*\w-\t) {$\bullet\ \IE$};
\end{tikzpicture}

%% file: fig-symmetrization.tex
\def\h{0.40cm} 
\def\w{1.00cm} 
\def\d{0.80cm} 

\begin{tikzpicture}
  [ctrl/.style = {circle, draw, solid, fill, inner sep = 0mm, minimum size = 1.1mm},
   gate/.style = {draw, fill = white,    thick, minimum width = \d, rectangle},
   orac/.style = {draw, fill = black!15, thick, minimum width = \d, rectangle},
   draw = black, label distance = -1pt
  ]

  \newcommand{\gate}[6]{
    \draw (#1*\w,#2*\h) node (#5) [#4, minimum height = #3*\h] {#6};
  }

  \newcommand{\cgate}[8]{
    \gate{#1}{#2}{#3}{#4}{#5}{#6};
    \draw (#1*\w,#7*\h) node (c#5) [ctrl, label = above:#8] {};
    \draw [-, thick] (c#5) to (#5);
  }

  \foreach \i in {0,...,2}
    \draw [-, thick] (0,\i*\h) +(-\d,0) to +(9*\w+\d,0);

  \foreach \i in {0,1,2}
    \gate{4*\i}{1}{3}{gate}{U\i}{$U_{\i}$};

  \gate{2}{1}{3}{orac}{O1}{$O_{\bv{x},u}$};
  \gate{6}{1}{3}{orac}{O2}{$O_{\bv{x},u}\ct$};

  \begin{scope}[black!60, densely dashed]
    \draw (-\w,3.5*\h) +(3pt,-2.4pt) node [scale = 0.7] {$\displaystyle\twirlingsquare{\ket{\bv{y}}\ket{v}} \;\; \bigg\{$};
    \foreach \i in {3,4}
      \draw [-, thick] (0,\i*\h) +(\w-\d/2,0) to +(9*\w+\d,0);
    \cgate{1}{1}{3}{gate}{v1}{$v$}{3}{$v$};
    \cgate{3}{1}{3}{gate}{V1}{$V_{\bv{y}}$}{4}{$\bv{y}$};
    \cgate{5}{1}{3}{gate}{V2}{$V_{\bv{y}}\ct$}{4}{$\bv{y}$};
    \cgate{7}{1}{3}{gate}{v2}{$v\ct$}{3}{$v$};
    \cgate{9}{1}{3}{gate}{V2}{$V_{\bv{y}}\ct$}{4}{$\bv{y}$};
  \end{scope}

\end{tikzpicture}

%% file: fig-Ue.tex
\def\h{0.60cm}
\def\w{1.20cm}

\begin{tikzpicture}
  [dot/.style = {circle, draw = black, fill = black, inner sep = 0mm, minimum size = 1.1mm},
   gate/.style = {draw = black, fill = white,    thick, minimum width = 0.5*\w, rectangle},
   orac/.style = {draw = black, fill = black!15, thick, minimum width = 0.5*\w, rectangle}
  ]

  \def\I{-0.7*\w}
  \def\W{ 1.7*\w}

  \draw (\I,+\h) node (q1) {};
  \draw (\I,  0) node (q2) {};
  \draw (\I,-\h) node (q3) {};

  \draw (q1) +(-0.6,0) node [right] {$\0{d}$};
  \draw (q2) +(-0.6,0) node [right] {$\0{n}$};
  \draw (q3) +(-0.6,0) node [right] {$\0{2}$};
  
  \draw [-, thick] (q1) to (\W,+\h);
  \draw [-, thick] (q2) to (\W,  0);
  \draw [-, thick] (q3) to (\W,-\h);

  \newcommand{\gate}[6]{
    \draw (#1*\w,#2*\h) node (#5) [#4, minimum height = #3*\h] {#6};
  }

  \gate{0}{0.5}{2}{orac}{O}{$O$};
  \gate{1}{-1 }{1}{gate}{R}{$\Rotk$};

  \draw [-, thick] (\w,0) to (R);
  \draw (\w,0) node [dot, label = above:$k$] {};
\end{tikzpicture}

%% file: fig-checking.tex
\def\h{0.60cm}
\def\w{1.20cm}

\begin{tikzpicture}
  [dot/.style = {circle, draw = black, fill = black, inner sep = 0mm, minimum size = 1.1mm},
   crc/.style = {circle, draw = black, fill = white, inner sep = 0mm, minimum size = 1.1mm},
   gate/.style = {draw = black, fill = white,    thick, minimum width = 0.5*\w, rectangle},
   orac/.style = {draw = black, fill = black!15, thick, minimum width = 0.5*\w, rectangle}
  ]

  \def\I{-0.7*\w}
  \def\W{ 3.7*\w}

  \draw (\I,+\h) node (q1) {};
  \draw (\I,  0) node (q2) {};

  \draw (q1) +(-0.6,0) node [right] {$\0{2}$};
  \draw (q2) +(-0.6,0) node [right] {$\0{n}$};
  
  \draw [-, thick] (q1) to (\W,+\h);
  \draw [-, thick] (q2) to (\W,  0);

  \newcommand{\gate}[6]{
    \draw (#1*\w,#2*\h) node (#5) [#4, minimum height = #3*\h] {#6};
  }

  \gate{0}{1}{1}{gate}{H1}{$H$};
  \gate{3}{1}{1}{gate}{H2}{$H$};
  \gate{1}{0}{1}{gate}{V}{$V$};
  \gate{2}{0}{1}{gate}{W}{$W$};

  \draw (1*\w,\h) node (C1) [dot] {};
  \draw (2*\w,\h) node (C2) [crc] {};

  \draw [-, thick] (C1) to (V);
  \draw [-, thick] (C2) to (W);
\end{tikzpicture}

%% file: qrs-itcs.bbl
\begin{thebibliography}{10}

\bibitem{Aaronson:2009}
S.~Aaronson.
\newblock Quantum copy-protection and quantum money.
\newblock In {\em Proceedings of the 24th Annual IEEE Conference on
  Computational Complexity (CCC'09)}, pages 229--242. IEEE Computer Society,
  2009.
\newblock \href {http://dx.doi.org/10.1109/CCC.2009.42}
  {\path{doi:10.1109/CCC.2009.42}}.

\bibitem{AD:2010}
S.~Aaronson and A.~Drucker.
\newblock A full characterization of quantum advice.
\newblock In {\em Proceedings of the 42nd Annual ACM Symposium on Theory of
  Computing (STOC'10)}, pages 131--140. ACM, 2010.
\newblock \href {http://arxiv.org/abs/1004.0377} {\path{arXiv:1004.0377}},
  \href {http://dx.doi.org/10.1145/1806689.1806710}
  {\path{doi:10.1145/1806689.1806710}}.

\bibitem{AR:2005}
D.~Aharonov and O.~Regev.
\newblock Lattice problems in {NP} $\cap$ co{NP}.
\newblock {\em Journal of the ACM}, 52(5):749--765, 2005.
\newblock (Earlier version in FOCS'04).
\newblock \href {http://dx.doi.org/10.1145/1089023.1089025}
  {\path{doi:10.1145/1089023.1089025}}.

\bibitem{aharonov03}
D.~Aharonov and A.~Ta-Shma.
\newblock Adiabatic quantum state generation and statistical zero knowledge.
\newblock In {\em Proceedings of the 35th Annual ACM Symposium on Theory of
  Computing (STOC'03)}, pages 20--29. ACM, 2003.
\newblock \href {http://arxiv.org/abs/quant-ph/0301023}
  {\path{arXiv:quant-ph/0301023}}, \href
  {http://dx.doi.org/10.1145/780542.780546} {\path{doi:10.1145/780542.780546}}.

\bibitem{Amb00}
A.~Ambainis.
\newblock Quantum lower bounds by quantum arguments.
\newblock In {\em Proceedings of the 32nd Annual ACM Symposium on Theory of
  Computing (STOC'00)}, pages 636--643. ACM, 2000.
\newblock \href {http://arxiv.org/abs/quant-ph/0002066}
  {\path{arXiv:quant-ph/0002066}}, \href
  {http://dx.doi.org/10.1145/335305.335394} {\path{doi:10.1145/335305.335394}}.

\bibitem{Ambainis2010}
A.~Ambainis.
\newblock {Variable time amplitude amplification and a faster quantum algorithm
  for solving systems of linear equations}.
\newblock 2010.
\newblock \href {http://arxiv.org/abs/1010.4458} {\path{arXiv:1010.4458}}.

\bibitem{AMRR:2011}
A.~Ambainis, L.~Magnin, M.~Roetteler, and J.~Roland.
\newblock Symmetry-assisted adversaries for quantum state generation.
\newblock In {\em Proceedings of the 26th Annual IEEE Conference on
  Computational Complexity (CCC'11)}, pages 167--177. IEEE Computer Society,
  2011.
\newblock \href {http://arxiv.org/abs/1012.2112} {\path{arXiv:1012.2112}},
  \href {http://dx.doi.org/10.1109/CCC.2011.24}
  {\path{doi:10.1109/CCC.2011.24}}.

\bibitem{bbbv97}
C.~H. Bennett, E.~Bernstein, G.~Brassard, and U.~Vazirani.
\newblock Strengths and weaknesses of quantum computing.
\newblock {\em SIAM Journal on Computing}, 26(5):1510--1523, 1997.
\newblock \href {http://arxiv.org/abs/quant-ph/9701001}
  {\path{arXiv:quant-ph/9701001}}, \href
  {http://dx.doi.org/10.1137/S0097539796300933}
  {\path{doi:10.1137/S0097539796300933}}.

\bibitem{Berry2006}
D.~W. Berry, G.~Ahokas, R.~Cleve, and B.~C. Sanders.
\newblock {Efficient quantum algorithms for simulating sparse Hamiltonians}.
\newblock {\em Communications in Mathematical Physics}, 270(2):9, 2005.
\newblock \href {http://arxiv.org/abs/quant-ph/0508139}
  {\path{arXiv:quant-ph/0508139}}, \href
  {http://dx.doi.org/10.1007/s00220-006-0150-x}
  {\path{doi:10.1007/s00220-006-0150-x}}.

\bibitem{BKS:2009}
S.~Boixo, E.~Knill, and R.~D. Somma.
\newblock Eigenpath traversal by phase randomization.
\newblock {\em Quantum Information and Computation}, 9(9,10):833--855, 2009.
\newblock \href {http://arxiv.org/abs/0903.1652} {\path{arXiv:0903.1652}}.

\bibitem{BoyerBHT98}
M.~Boyer, G.~Brassard, P.~H\o~yer, and A.~Tapp.
\newblock {Tight bounds on quantum searching}.
\newblock {\em Fortschritte der Physik}, 46(4-5):493--505, 1998.
\newblock \href
  {http://dx.doi.org/10.1002/(SICI)1521-3978(199806)46:4/5<493::AID-PROP493>3.0.CO;2-P}
  {\path{doi:10.1002/(SICI)1521-3978(199806)46:4/5<493::AID-PROP493>3.0.CO;2-P}}.

\bibitem{AmplitudeAmplification}
G.~Brassard, P.~H\o{}yer, M.~Mosca, and A.~Tapp.
\newblock Quantum amplitude amplification and estimation.
\newblock 2000.
\newblock \href {http://arxiv.org/abs/quant-ph/0005055}
  {\path{arXiv:quant-ph/0005055}}.

\bibitem{BCWW:2001}
H.~Buhrman, R.~Cleve, J.~Watrous, and R.~de~Wolf.
\newblock Quantum fingerprinting.
\newblock {\em Physical Review Letters}, 87(16):167902, 2001.
\newblock \href {http://arxiv.org/abs/quant-ph/0102001}
  {\path{arXiv:quant-ph/0102001}}, \href
  {http://dx.doi.org/10.1103/PhysRevLett.87.167902}
  {\path{doi:10.1103/PhysRevLett.87.167902}}.

\bibitem{BuhrmanDeWolf02querysurvey}
H.~Buhrman and R.~de~Wolf.
\newblock Complexity measures and decision tree complexity: A survey.
\newblock {\em Theoretical Computer Science}, 288(1):21--43, 2002.
\newblock \href {http://dx.doi.org/10.1016/S0304-3975(01)00144-X}
  {\path{doi:10.1016/S0304-3975(01)00144-X}}.

\bibitem{Childs2009a}
A.~M. Childs.
\newblock {On the relationship between continuous- and discrete-time quantum
  walk}.
\newblock {\em Communications in Mathematical Physics}, 294(2):22, 2008.
\newblock \href {http://arxiv.org/abs/0810.0312} {\path{arXiv:0810.0312}},
  \href {http://dx.doi.org/10.1007/s00220-009-0930-1}
  {\path{doi:10.1007/s00220-009-0930-1}}.

\bibitem{Childs2011}
A.~M. Childs and R.~Kothari.
\newblock {Simulating Sparse Hamiltonians with Star Decompositions}.
\newblock In {\em Theory of Quantum Computation, Communication, and
  Cryptography (TQC 2010)}, volume 6519 of {\em Lecture Notes in Computer
  Science}, page~11, Berlin, Heidelberg, 2011. Springer.
\newblock \href {http://arxiv.org/abs/1003.3683} {\path{arXiv:1003.3683}},
  \href {http://dx.doi.org/10.1007/978-3-642-18073-6}
  {\path{doi:10.1007/978-3-642-18073-6}}.

\bibitem{CEMM98}
R.~Cleve, A.~Ekert, C.~Macchiavello, and M.~Mosca.
\newblock {Quantum Algorithms Revisited}.
\newblock {\em Proceedings of the Royal Society A: Mathematical, Physical and
  Engineering Sciences}, 454(1969):18, 1997.
\newblock \href {http://arxiv.org/abs/quant-ph/9708016}
  {\path{arXiv:quant-ph/9708016}}, \href
  {http://dx.doi.org/10.1098/rspa.1998.0164}
  {\path{doi:10.1098/rspa.1998.0164}}.

\bibitem{vDHI:2003}
W.~Dam, S.~Hallgren, and L.~Ip.
\newblock Quantum algorithms for some hidden shift problems.
\newblock In {\em {Proceedings of the 14th Annual ACM-SIAM Symposium on
  Discrete Algorithms (SODA'03)}}, pages 489--498, 2003.
\newblock \href {http://arxiv.org/abs/quant-ph/0211140}
  {\path{arXiv:quant-ph/0211140}}.

\bibitem{DeWolf:2008}
R.~de~Wolf.
\newblock A brief introduction to {F}ourier analysis on the {B}oolean cube.
\newblock {\em Theory of Computing Library -- Graduate Surveys}, 1:1--20, 2008.
\newblock \href {http://dx.doi.org/10.4086/toc.gs.2008.001}
  {\path{doi:10.4086/toc.gs.2008.001}}.

\bibitem{devroye}
L.~Devroye.
\newblock {\em Non-uniform random variate generation}.
\newblock Springer, New York, 1986.

\bibitem{FGH+:2010}
E.~Farhi, D.~Gosset, A.~Hassidim, A.~Lutomirski, and P.~Shor.
\newblock Quantum money from knots.
\newblock In {\em Proceedings of the 3rd Innovations in Theoretical Computer
  Science (ITCS'12) conference}, 2012.
\newblock To appear.
\newblock \href {http://arxiv.org/abs/1004.5127} {\path{arXiv:1004.5127}}.

\bibitem{Friedl2003}
K.~Friedl, G.~Ivanyos, F.~Magniez, M.~Santha, and P.~Sen.
\newblock {Hidden Translation and Orbit Coset in Quantum Computing}.
\newblock In {\em Proceedings of the 35fth Annual ACM Symposium on Theory of
  Computing (STOC'03)}, pages 1--9. ACM, 2002.
\newblock \href {http://arxiv.org/abs/quant-ph/0211091}
  {\path{arXiv:quant-ph/0211091}}, \href
  {http://dx.doi.org/10.1145/780542.780544} {\path{doi:10.1145/780542.780544}}.

\bibitem{Grover:96}
L.~K. Grover.
\newblock A fast quantum mechanical algorithm for database search.
\newblock In {\em Proceedings of the 28th Annual ACM Symposium on Theory of
  Computing (STOC'96)}, pages 212--219, New York, 1996. ACM.
\newblock \href {http://arxiv.org/abs/quant-ph/9605043}
  {\path{arXiv:quant-ph/9605043}}, \href
  {http://dx.doi.org/10.1145/237814.237866} {\path{doi:10.1145/237814.237866}}.

\bibitem{Grover00}
L.~K. Grover.
\newblock Synthesis of quantum superpositions by quantum computation.
\newblock {\em Physical Review Letters}, 85(6):1334--1337, 2000.
\newblock \href {http://dx.doi.org/10.1103/PhysRevLett.85.1334}
  {\path{doi:10.1103/PhysRevLett.85.1334}}.

\bibitem{Grover2002}
L.~K. Grover and T.~Rudolph.
\newblock {Creating superpositions that correspond to efficiently integrable
  probability distributions}.
\newblock 2002.
\newblock \href {http://arxiv.org/abs/arxiv:0208112}
  {\path{arXiv:arxiv:0208112}}.

\bibitem{HHL:09}
A.~W. Harrow, A.~Hassidim, and S.~Lloyd.
\newblock Quantum algorithm for linear systems of equations.
\newblock {\em Physical Review Letters}, 103(15):150502, 2009.
\newblock \href {http://arxiv.org/abs/0811.3171} {\path{arXiv:0811.3171}},
  \href {http://dx.doi.org/10.1103/PhysRevLett.103.150502}
  {\path{doi:10.1103/PhysRevLett.103.150502}}.

\bibitem{NegativeWeights}
P.~H\o{}yer, T.~Lee, and R.~\v{S}palek.
\newblock Negative weights make adversaries stronger.
\newblock In {\em Proceedings of the 39th Annual ACM Symposium on Theory of
  Computing (STOC'07)}, pages 526--535. ACM, 2007.
\newblock \href {http://arxiv.org/abs/quant-ph/0611054}
  {\path{arXiv:quant-ph/0611054}}, \href
  {http://dx.doi.org/10.1145/1250790.1250867}
  {\path{doi:10.1145/1250790.1250867}}.

\bibitem{HMW03}
P.~H{\o}yer, M.~Mosca, and R.~de~Wolf.
\newblock Quantum search on bounded-error inputs.
\newblock In {\em Proceedings of the 30th International Colloquium on Automata,
  Languages and Programming (ICALP'03)}, volume 2719 of {\em Lecture Notes in
  Computer Science}, pages 291--299. Springer, 2003.
\newblock \href {http://arxiv.org/abs/quant-ph/0304052}
  {\path{arXiv:quant-ph/0304052}}, \href
  {http://dx.doi.org/10.1007/3-540-45061-0_25}
  {\path{doi:10.1007/3-540-45061-0_25}}.

\bibitem{Ivanyos:2008}
G.~Ivanyos.
\newblock On solving systems of random linear disequations.
\newblock {\em Quantum Information and Computation}, 8(6\&7):579--594, 2008.
\newblock URL: \url{http://www.rintonpress.com/journals/qiconline.html#v8n67},
  \href {http://arxiv.org/abs/0704.2988} {\path{arXiv:0704.2988}}.

\bibitem{KKR:2006}
J.~Kempe, A.~Kitaev, and O.~Regev.
\newblock The complexity of the local {H}amiltonian problem.
\newblock {\em SIAM Journal on Computing}, 35(5):1070--1097, 2006.
\newblock \href {http://arxiv.org/abs/quant-ph/0406180}
  {\path{arXiv:quant-ph/0406180}}, \href
  {http://dx.doi.org/10.1137/S0097539704445226}
  {\path{doi:10.1137/S0097539704445226}}.

\bibitem{Kitaev1995}
A.~Kitaev.
\newblock {Quantum measurements and the Abelian Stabilizer Problem}.
\newblock 1995.
\newblock \href {http://arxiv.org/abs/quant-ph/9511026}
  {\path{arXiv:quant-ph/9511026}}.

\bibitem{KW:2009}
A.~Kitaev and W.~A. Webb.
\newblock Wavefunction preparation and resampling using a quantum computer.
\newblock 2008.
\newblock \href {http://arxiv.org/abs/0801.0342} {\path{arXiv:0801.0342}}.

\bibitem{KST:93}
J.~K{\"o}bler, U.~Sch{\"o}ning, and J.~Toran.
\newblock {\em The Graph Isomorphism Problem: Its Structural Complexity}.
\newblock Progress in Theoretical Computer Science. Birkh{\"a}user Boston,
  1993.

\bibitem{LMRSS11}
T.~Lee, R.~Mittal, B.~W. Reichardt, R.~{\v{S}}palek, and M.~Szegedy.
\newblock Quantum query complexity of state conversion.
\newblock In {\em Proceedings of the 52nd Annual IEEE Symposium on Foundations
  of Computer Science (FOCS'11)}, 2011.
\newblock To appear.
\newblock \href {http://arxiv.org/abs/1011.3020} {\path{arXiv:1011.3020}}.

\bibitem{Letac75}
G.~Letac.
\newblock On building random variables of a given distribution.
\newblock {\em The Annals of Probability}, 3(2):298--306, 1975.
\newblock \href {http://dx.doi.org/10.1214/aop/1176996400}
  {\path{doi:10.1214/aop/1176996400}}.

\bibitem{MW05}
C.~Marriott and J.~Watrous.
\newblock Quantum {A}rthur--{M}erlin games.
\newblock {\em Computational Complexity}, 14(2):122--152, 2005.
\newblock \href {http://arxiv.org/abs/cs/0506068} {\path{arXiv:cs/0506068}},
  \href {http://dx.doi.org/10.1007/s00037-005-0194-x}
  {\path{doi:10.1007/s00037-005-0194-x}}.

\bibitem{metropolis53}
N.~Metropolis, A.~W. Rosenbluth, M.~N. Rosenbluth, A.~H. Teller, and E.~Teller.
\newblock Equation of state calculations by fast computing machines.
\newblock {\em The Journal of Chemical Physics}, 21(6):1087, 1953.
\newblock \href {http://dx.doi.org/10.1063/1.1699114}
  {\path{doi:10.1063/1.1699114}}.

\bibitem{NWZ09}
D.~Nagaj, P.~Wocjan, and Y.~Zhang.
\newblock Fast amplification of {QMA}.
\newblock {\em Quantum Information and Computation}, 9(11\&12):1053--1068,
  2009.
\newblock URL:
  \url{http://www.rintonpress.com/journals/qiconline.html#v9n1112}, \href
  {http://arxiv.org/abs/0904.1549} {\path{arXiv:0904.1549}}.

\bibitem{Regev:2004}
O.~Regev.
\newblock Quantum computation and lattice problems.
\newblock {\em SIAM Journal on Computing}, 33(2):738--760, 2004.
\newblock \href {http://arxiv.org/abs/cs/0304005} {\path{arXiv:cs/0304005}},
  \href {http://dx.doi.org/10.1137/S0097539703440678}
  {\path{doi:10.1137/S0097539703440678}}.

\bibitem{Reichardt:2009}
B.~W. Reichardt.
\newblock {Span Programs and Quantum Query Complexity: The General Adversary
  Bound Is Nearly Tight for Every Boolean Function}.
\newblock In {\em Proceedings of the 50th Annual IEEE Symposium on Foundations
  of Computer Science (FOCS'09)}, pages 544--551. IEEE Computer Society Press,
  2009.
\newblock \href {http://arxiv.org/abs/0904.2759} {\path{arXiv:0904.2759}},
  \href {http://dx.doi.org/10.1109/FOCS.2009.55}
  {\path{doi:10.1109/FOCS.2009.55}}.

\bibitem{ReichardtReflections}
B.~W. Reichardt.
\newblock Reflections for quantum query algorithms.
\newblock In {\em Proceedings of the 22nd Annual ACM-SIAM Symposium on Discrete
  Algorithms (SODA'11)}, pages 560--569, 2011.
\newblock \href {http://arxiv.org/abs/1005.1601} {\path{arXiv:1005.1601}}.

\bibitem{Roetteler:2010}
M.~R{\"o}tteler.
\newblock Quantum algorithms for highly non-linear {B}oolean functions.
\newblock In {\em Proceedings of the 21st Annual ACM-SIAM Symposium on Discrete
  Algorithms (SODA'10)}, pages 448--457, 2010.
\newblock \href {http://arxiv.org/abs/0811.3208} {\path{arXiv:0811.3208}}.

\bibitem{STV11}
M.~Schwarz, K.~Temme, and F.~Verstraete.
\newblock Contracting tensor networks and preparing {PEPS} on a quantum
  computer.
\newblock arxiv:1104.1410, 2011.
\newblock \href {http://arxiv.org/abs/1104.1410} {\path{arXiv:1104.1410}}.

\bibitem{Sheridan2009}
L.~Sheridan, D.~Maslov, and M.~Mosca.
\newblock {Approximating fractional time quantum evolution}.
\newblock {\em Journal of Physics A}, 42(18):185302, 2009.
\newblock \href {http://arxiv.org/abs/0810.3843} {\path{arXiv:0810.3843}},
  \href {http://dx.doi.org/10.1088/1751-8113/42/18/185302}
  {\path{doi:10.1088/1751-8113/42/18/185302}}.

\bibitem{SBBK:2008}
R.~D. Somma, S.~Boixo, H.~Barnum, and E.~Knill.
\newblock Quantum simulations of classical annealing processes.
\newblock {\em Physical Review Letters}, 101:130504, 2008.
\newblock \href {http://arxiv.org/abs/0804.1571} {\path{arXiv:0804.1571}},
  \href {http://dx.doi.org/10.1103/PhysRevLett.101.130504}
  {\path{doi:10.1103/PhysRevLett.101.130504}}.

\bibitem{TOV+:2009}
K.~Temme, T.~J. Osborne, K.~G. Vollbrecht, D.~Poulin, and F.~Verstraete.
\newblock Quantum {M}etropolis sampling.
\newblock {\em Nature}, 471(7336):87--90, 2011.
\newblock \href {http://arxiv.org/abs/0911.3635} {\path{arXiv:0911.3635}},
  \href {http://dx.doi.org/10.1038/nature09770}
  {\path{doi:10.1038/nature09770}}.

\bibitem{VB:96}
L.~Vandenberghe and S.~Boyd.
\newblock {Semidefinite programming}.
\newblock {\em SIAM Review}, 38(1):49, 1996.

\bibitem{Vazirani:98}
U.~Vazirani.
\newblock On the power of quantum computation.
\newblock {\em Philosophical Transactions: Mathematical, Physical and
  Engineering Sciences}, 356(1743):1759--1768, 1998.
\newblock URL: \url{http://www.jstor.org/stable/55010}.

\bibitem{vonNeumann51}
J.~von Neumann.
\newblock Various techniques used in connection with random digits.
\newblock {\em National Bureau of Standards, Applied Math Series}, 12:36--38,
  1951.

\bibitem{Watrous:2000}
J.~Watrous.
\newblock Succinct quantum proofs for properties of finite groups.
\newblock In {\em Proceedings of the 41st Annual IEEE Symposium on Foundations
  of Computer Science (FOCS'00)}, pages 537--546. IEEE Computer Society, 2000.
\newblock \href {http://arxiv.org/abs/cs/0009002} {\path{arXiv:cs/0009002}},
  \href {http://dx.doi.org/10.1109/SFCS.2000.892141}
  {\path{doi:10.1109/SFCS.2000.892141}}.

\bibitem{Watrous2001}
J.~Watrous.
\newblock {Quantum algorithms for solvable groups}.
\newblock In {\em Proceedings of the 33rd Annual ACM Symposium on Theory of
  Computing (STOC'01)}, pages 60--67. ACM, 2001.
\newblock \href {http://arxiv.org/abs/quant-ph/0011023}
  {\path{arXiv:quant-ph/0011023}}, \href
  {http://dx.doi.org/10.1145/380752.380759} {\path{doi:10.1145/380752.380759}}.

\bibitem{YA:2010}
M.-H. Yung and A.~Aspuru-Guzik.
\newblock A quantum-quantum {M}etropolis algorithm.
\newblock 2010.
\newblock \href {http://arxiv.org/abs/1011.1468} {\path{arXiv:1011.1468}}.

\end{thebibliography}
